\documentclass[trackchanges,twocolumn,twocolappendix]{aastex701}
\usepackage{array}
\usepackage{booktabs}
\usepackage{xcolor}
\usepackage{graphicx} 
\usepackage{caption}  
\usepackage{graphicx}
\usepackage{float}

\hypersetup{linkcolor=red,citecolor=blue,filecolor=cyan,urlcolor=purple}

\begin{document}

\title{Compiling the largest multiwavelength database of known changing-look AGN: Prospects in the Rubin era}

\correspondingauthor{Swayamtrupta Panda}
\email{swayamtrupta.panda@noirlab.edu}

\author[orcid=0009-0000-1806-5302, sname='Camus']{Mariangella Camus}
\affiliation{Pontificia Universidad Católica de Valparaíso, Chile}
\email{mariangella.camus@pucv.cl}

\author[orcid=0000-0002-5854-7426, sname='Panda']{Swayamtrupta Panda}\thanks{Gemini Science Fellow}
\affiliation{International Gemini Observatory/NSF NOIRLab, Casilla 603, La Serena, Chile}
\email{swayamtrupta.panda@noirlab.edu}

\begin{abstract}
Changing-look active galactic nuclei (CL-AGN) provide a direct way to study accretion and spectral-state changes on timescales of months to years. Still, the known population remains scattered across heterogeneous literature samples. We present a compilation of a homogenized catalog of known CL-AGN and related candidates, designed as a traceable reference database for time-domain and multiwavelength studies. After coordinate homogenization, duplicate removal, and validation, the optical parent catalog contains 1,438 unique sources, including 987 spectroscopically confirmed CL-AGN and 451 candidate or photometrically selected systems. We augment the catalog with ancillary information from radio/mm, infrared, ultraviolet, X-ray, and gamma-ray data sets using \texttt{nway} and survey-specific quality criteria. The resulting database includes 188 sources in the broad radio layer, 9 RFC/VLBI compact-core matches, 16 sources with ALMA archival coverage, 1,353 infrared counterparts, 918 GALEX ultraviolet matches, and 576 X-ray counterparts. We also identify 9 Fermi-LAT candidate associations. Overall, 1,385 sources have at least one non-optical counterpart, and 75 have coverage across the four main ancillary regimes: radio, infrared, ultraviolet, and X-ray. As a first application, we identify 70 confirmed CL-AGN within the LSST Wide--Fast--Deep footprint and 5 within the Rubin Deep Drilling Fields, and provide the up-to-date, $\sim$116-day \textit{ugrizy} Rubin lightcurve for SDSS J095902.76+021906.3 (z=0.34579) and discuss the early data inferences. This catalog is envisioned to be community-driven and provides a dynamic, expandable resource for target selection, population comparisons, and future Rubin-era CL-AGN studies.

\end{abstract}

\section{Introduction} \label{sec:intro} Active galactic nuclei (AGN) are powered by accretion onto supermassive black holes, and their observed emission carries information about the structure and time-dependent behavior of the central engine. Although AGN were traditionally described through relatively stable observational classes, long-term monitoring has shown that a small but important fraction of them can undergo dramatic changes on timescales of months to years. Among the most striking examples are changing-look active galactic nuclei (CL-AGN), whose spectra and/or photometric properties change enough to alter their observational classification \citep{ricci2023changing}. These transitions may involve the appearance or disappearance of broad emission lines, strong continuum variability, or changes between different Seyfert-like spectral types \citep{lamassa2015discovery,Runnoe2016MNRAS.455.1691R, Macleod2016MNRAS.457..389M, macleod2019changing,zeltyn2024exploring}. Because these changes usually occur on humanly accessible timescales \citep{lamassa2015discovery, Stern2018ApJ...864...27S, Trakhtenbrot2019ApJ...883...94T, Zeltyn2022ApJ...939L..16Z}, CL-AGN offer a direct way to study the connection between accretion variability, broad-line region response, obscuration, and the evolving structure of AGN.\\

The physical origin of changing-look behavior is still an active area of investigation. In some sources, the observed transition is likely connected to intrinsic changes in the accretion flow, where a change in the ionizing continuum modifies the broad emission lines and the optical/UV continuum \citep{Noda2018MNRAS.480.3898N, Sniegowska2020A&A...641A.167S, Panda2022AN....34310091P}. In other cases, variable obscuration, changes in the line-of-sight column density, or a combination of accretion and geometric effects may contribute to the observed transformation \citep{lamassa2015discovery, Stern2018ApJ...864...27S, Gaskell2018MNRAS.478.1660G, Wang2019ApJ...887...15W, Yang2019ApJ...885..110Y}. Distinguishing between these scenarios requires well-characterized samples with reliable coordinates, redshifts, classifications, multi-epoch information, and multiwavelength context. Individual CL-AGN are therefore valuable, but a homogeneous catalog is essential for understanding the population as a whole.\\ 

Over the last decade, the number of reported CL-AGN has increased rapidly due to repeated spectroscopy, long-baseline photometric monitoring, and systematic searches in large surveys \citep{macleod2019changing, hon2022skymapper, zeltyn2024exploring, guo2025changing}. However, the known population remains highly fragmented across the literature. Different works adopt different selection methods, naming conventions, coordinate formats, spectral classifications, and criteria for defining a changing-look event. Some samples focus on spectroscopically confirmed transitions, while others include photometrically selected candidates or changing-state AGN identified through variability and follow-up observations \citep[see e.g.,][]{SanchezSaez2021AJ....162..206S, LopezNavas2023MNRAS.524..188L, ShuWang2024ApJ...966..128W}. As a result, the current literature provides a rich but heterogeneous view of the CL-AGN population. This fragmentation creates practical limitations for future time-domain studies. Without a unified catalog, it is difficult to identify duplicate sources, compare properties across samples, determine which objects have multiwavelength coverage, or select targets for follow-up observations. This is especially important in the context of upcoming and ongoing wide-field surveys, where known CL-AGN can serve both as test cases for variability studies and as training examples for identifying new candidates. A compiled and homogenized catalog is therefore not only a bookkeeping exercise, but a necessary step toward building a more complete observational framework for CL-AGN.\\ 

The need for such a compilation is particularly relevant for the Southern Hemisphere. Several major CL-AGN searches have been based on northern or equatorial surveys \citep[see e.g.,][]{macleod2019changing, zeltyn2024exploring, guo2025changing}, while the southern sky remains less uniformly organized from the perspective of known changing-look sources. This is important because the Vera C. Rubin Observatory Legacy Survey of Space and Time (LSST) is set to provide an unprecedented wide-field, multi-epoch view of the southern sky \citep{Ivezic2019}. Rubin will repeatedly observe millions of variable sources, making it possible to identify new changing-look candidates, monitor known CL-AGN, and connect optical variability with archival data at other wavelengths \citep[see][for a recent review]{2026arXiv260121769P}. To take full advantage of this opportunity, it is necessary to know which previously reported CL-AGN fall within the Rubin footprint and what ancillary information is already available for them.\\ 

In this work, we present a compiled catalog of known CL-AGN collected from the literature and homogenized into a common and accessible format. The catalog includes source names, sky positions, redshifts where available, literature references (source publication and associated bibliography), and classification information when reported. We also perform a multiwavelength cross-match to connect the CL-AGN sample with existing photometric, spectroscopic, infrared, X-ray, and radio information. This cross-match is designed to make the catalog useful not only as a static list of objects, but also as a practical database for selecting targets, comparing populations, and planning future follow-up.\\ 

As a first demonstration of the scientific use of the catalog, we examine the overlap between known CL-AGN and Rubin observing regions. In particular, we identify sources located within the Wide-Fast-Deep footprint and within the Rubin Deep Drilling Fields, where high-cadence observations and extensive archival coverage provide especially favorable conditions for future variability studies\footnote{A preliminary version of this experiment can be found in \citet{camus2026known}.}. Fields such as COSMOS \citep{COSMOS2007ApJS..172....1S, COSMOS2010ApJ...716..348B, COSMOS2016ApJS..224...24L} and XMM-LSS \citep{XMMLSS2004JCAP...09..011P, XMMLSS2016A&A...592A...1P} are especially valuable because they combine repeated optical monitoring with deep multiwavelength legacy data, making them natural laboratories for studying the physical drivers of changing-look activity.\\ 

The goal of this paper is therefore twofold. First, we present a unified and expandable catalog of known CL-AGN, documenting how the sample was assembled, homogenized, and cross-matched with external surveys. This is done keeping in mind that our methodology can be easily reproducible. Second, we show how this catalog can be used in practice through a Rubin-focused application. This first paper is intentionally focused on the construction, validation, and immediate utility of the catalog, rather than on a detailed physical interpretation of every source. More complex population studies, including detailed light-curve modeling, spectroscopic follow-up, and machine-learning classification of new candidates, are ongoing and will be developed in subsequent works.\\

This paper is organized as follows. 
In Section~\ref{sec:data}, we describe the literature samples used to construct the master CL-AGN catalog and the homogenization procedure. 
In Section~\ref{sec:rubin}, we describe the Rubin footprint comparison and the identification of sources in high-cadence fields. 
In Section~\ref{sec:crossmatch}, we present the multiwavelength cross-match and summarize the information extracted from each external catalog. 
In Section~\ref{sec:mw_results}, we present the main catalog statistics and multiwavelength coverage. 
In Sections~\ref{sec:database} and~\ref{sec:discussion}, we describe the database structure, its limitations, and future extensions, and summarize our results in Section~\ref{sec:summary}.\\

\section{Parent sample and catalog construction}
\label{sec:data}

The starting point of this work is a literature-based catalog of known changing-look active galactic nuclei (CL-AGN) and closely related candidates. The goal was not to define a new selection function, but to build a clean and traceable reference sample from the sources that have already been reported in the literature. In this sense, the catalog should be viewed as a compiled database of known CL-AGN rather than as a statistically complete survey sample.\\

The parent sample was assembled from published CL-AGN searches, spectroscopic studies, and candidate compilations using the NASA Astrophysics Data System \citep{NASAADS2000A&AS..143...41K}\footnote{\url{https://ui.adsabs.harvard.edu/}}. We included sources identified through repeated spectroscopy, broad-line appearance or disappearance, optical variability, changing-state behavior, and follow-up observations. The input literature includes SDSS-based and spectroscopic searches \citep{yang2018discovery,macleod2019changing,zeltyn2024exploring,dong2025newly}, SkyMapper and 6dF-related searches \citep{hon2022skymapper,amrutha2024discovering}, Swift-BAT/BASS changing-look AGN \citep{temple2023bass}, DESI changing-look AGN samples \citep{guo2025changing,chen2026changing}, and recent CL-AGN compilations and interpretive samples \citep{panda2024changing}. Together, these works form the optical parent sample used throughout this paper.\\

The input samples are heterogeneous by construction. They were selected from different surveys, with different depths, cadences, wavelength coverage, spectroscopic baselines, and classification criteria. Some works report spectroscopically confirmed transitions, while others include photometrically selected candidates or changing-state AGN that require further confirmation. For this reason, we preserve the literature provenance of each object instead of forcing all sources into a single uniform selection scheme. This allows the catalog to retain the context in which each source was originally identified.\\

For each source, we collected the information needed for identification and later cross-matching: source name, right ascension (J2000), declination (J2000), redshift when available, literature reference, and the reported changing-look or changing-state classification. Coordinates were converted to decimal degrees and placed into a common format before any spatial comparison was performed. When possible, names were also homogenized, but the original identifiers and references were kept so that each entry remains traceable to the literature.\\

Because the same astrophysical source can appear in more than one publication, duplicate removal was a central step in the construction of the catalog. We first searched for repeated entries using positional cross-matching, and then manually inspected ambiguous cases. When multiple literature entries corresponded to the same source, we kept a single catalog entry and preserved all associated references. This approach avoids double-counting while keeping the full literature history of the object.\\

After homogenization, duplicate removal, and validation, the final optical parent catalog contains 1,438 unique sources. Of these, 987 are spectroscopically confirmed CL-AGN and 451 are photometric or candidate systems. We keep both subsets in the database because they serve different purposes: the confirmed sample provides the most reliable reference set, while the candidate sample is useful for follow-up, comparison with future time-domain searches, and identifying objects that may become confirmed with additional spectroscopy.\\

Each source is assigned a unique internal identifier and is stored with its main name, coordinates, classification information, redshift when available, and literature references (source publication and/or discovery paper, and associated bibliography). The catalog is therefore designed to be expandable: new sources can be added in future versions without losing the origin of previous entries or mixing confirmed sources with candidates. This is particularly important in the Rubin era, where the number of reported changing-look candidates is expected to increase rapidly. Based on the current expectations, the 10-year survey would provide us with 2$\times$10$^5$ - 1.5$\times$10$^6$ photometric changing-look AGN candidates (see Sec. \ref{sec:summary} for more details). \\

The literature samples used to construct the parent catalog were compiled from the works cited above. 
Because these input samples are heterogeneous and partially overlapping, the numbers reported in individual studies should not be interpreted as independent additions to the final catalog. 
The final number of unique sources is obtained only after the coordinate homogenization, validation, and duplicate-removal procedure described above.

\begin{table*}[htbp]
\centering
\caption{
Summary of the multiwavelength ancillary layers included in the CL-AGN database. 
Counts and fractions are computed relative to the full optical parent catalog of 1,438 sources. 
Rows are not mutually exclusive, because a single CL-AGN can be detected in multiple catalogs or wavelength regimes.
}
\label{tab:multiwavelength_simple_summary}
\resizebox{\textwidth}{!}{%
\begin{tabular}{llcc}
\hline
Wavelength regime & Catalog or layer & Number of sources & Fraction of parent sample \\
\hline
Radio & VLASS & 91 & 6.3\% \\
Radio & FIRST & 89 & 6.2\% \\
Radio & NVSS & 97 & 6.7\% \\
Radio & RACS & 82 & 5.7\% \\
Radio & SUMSS & 7 & 0.5\% \\
Radio & LoTSS & 32 & 2.2\% \\
Radio & Any radio survey & 160 & 11.1\% \\
Radio/VLBI & RFC/VLBI compact core & 9 & 0.6\% \\
Millimeter/submillimeter & ALMA archival coverage & 16 & 1.1\% \\
\hline
Infrared & AllWISE & 1183 & 82.3\% \\
Infrared & CatWISE2020 & 1232 & 85.7\% \\
Infrared & 2MASS PSC & 489 & 34.0\% \\
Infrared & AKARI/IRC & 16 & 1.1\% \\
Infrared & AKARI/FIS & 19 & 1.3\% \\
Infrared & Any infrared catalog & 1353 & 94.1\% \\
\hline
Ultraviolet & GALEX best-match layer & 918 & 63.8\% \\
Ultraviolet & GALEX strong-match subset & 139 & 9.7\% \\
\hline
X-ray & ROSAT BSC & 97 & 6.7\% \\
X-ray & ROSAT FSC & 156 & 10.8\% \\
X-ray & ROSAT 2RXS & 256 & 17.8\% \\
X-ray & Any ROSAT layer & 285 & 19.8\% \\
X-ray & eRASS1 main catalog & 266 & 18.5\% \\
X-ray & eRASS1 supplementary catalog & 10 & 0.7\% \\
X-ray & eRASS1 hard catalog & 32 & 2.2\% \\
X-ray & Any eRASS1 layer & 276 & 19.2\% \\
X-ray & XMM-Newton 4XMM-DR13 & 135 & 9.4\% \\
X-ray & Chandra CSC & 89 & 6.2\% \\
X-ray & Any X-ray catalog & 576 & 40.1\% \\
\hline
Gamma-ray & Fermi-LAT candidate associations & 9 & 0.6\% \\
\hline
Summary & Any non-optical ancillary counterpart & 1385 & 96.3\% \\
Summary & Radio + infrared + ultraviolet + X-ray coverage & 75 & 5.2\% \\
\hline
\end{tabular}%
}
\end{table*}


\begin{figure*}[!t]
    \centering
    \includegraphics[width=0.95\linewidth]{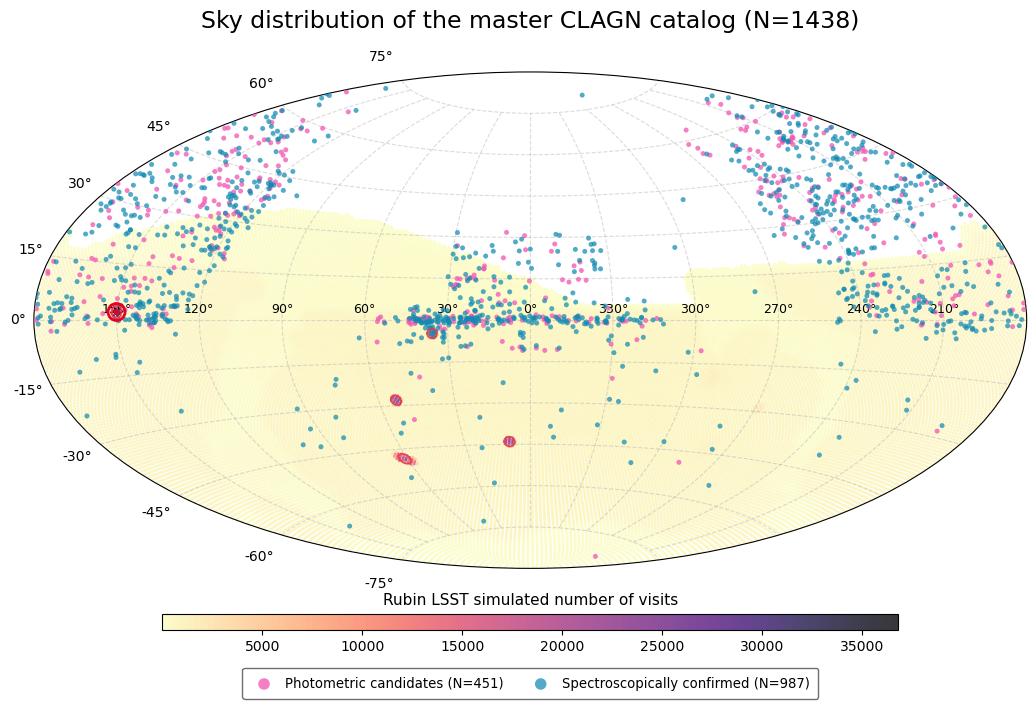}
   \caption{Sky distribution of the master CL-AGN catalog in equatorial coordinates. 
Pink points show candidate or photometrically selected systems, while cyan points show spectroscopically confirmed CL-AGN. The pale background shows the Rubin/LSST survey footprint, color-coded by the expected number of visits, and the red open outlines mark the Rubin Deep Drilling Fields. The catalog contains 1,438 unique sources after coordinate homogenization, literature consolidation, and duplicate removal, including 987 spectroscopically confirmed CL-AGN and 451 candidate systems. 
The sky distribution reflects the footprints and selection functions of the input literature samples rather than the intrinsic distribution of CL-AGN. We highlight the source, SDSS J095902.76+021906.3 (which is discussed in detail in Sec. \ref{sec:rubin_source}) with a bold red circle. 
}
    \label{fig:main_figure}
\end{figure*}


\section{Sky distribution and Rubin/LSST footprint}
\label{sec:rubin}

After constructing the optical parent catalog, we examined the sky distribution of the confirmed CL-AGN sample and its overlap with the Vera C. Rubin Observatory Legacy Survey of Space and Time (LSST). This step is useful for two reasons. First, it gives a simple view of the current observational biases in the known CL-AGN population. Second, it identifies previously reported CL-AGN that will be monitored by Rubin and can therefore serve as reference objects for future time-domain studies.

\subsection{Sky distribution and hemispheric bias}
\label{subsec:sky_distribution}

The known CL-AGN population is not distributed uniformly on the sky. This is expected because the parent catalog is compiled from the literature and therefore inherits the footprint, depth, cadence, and spectroscopic selection effects of the surveys from which the sources were originally identified. We therefore do not interpret the sky distribution as an intrinsic distribution of CL-AGN. Instead, we use it as a diagnostic of where the current observational coverage is strongest and where future surveys can improve the sample.\\

The sky distribution of the full optical parent catalog, separated into spectroscopically confirmed CL-AGN and candidate systems and shown together with the Rubin/LSST footprint and Deep Drilling Fields, is presented in Figure~\ref{fig:main_figure}.\\

A clear feature of the confirmed sample is the deficit of sources in the Southern Hemisphere. Only 21\% of the spectroscopically confirmed CL-AGN in the catalog are located below the Celestial Equator (i.e., with negative declinations). This fraction highlights the strong northern and equatorial bias of the current literature sample, which is largely driven by the historical availability of repeated spectroscopy and long-baseline monitoring in surveys such as SDSS and DESI. The southern deficit is important because it means that a large part of the sky remains underrepresented in the current census of known CL-AGN. This provides a natural motivation for coordinated southern follow-up, including optical monitoring and complementary radio facilities for AGN studies (Camus et al., in prep.).\\

This bias is especially relevant in the Rubin era. LSST will repeatedly observe the southern sky with high cadence and depth, providing an opportunity to identify new changing-look candidates and to monitor known sources with uniform optical light curves. A cleaned catalog of previously reported CL-AGN is therefore useful not only as a reference list, but also as a way to quantify which known objects already fall in Rubin-observed regions and which regions remain poorly represented.\\

\subsection{Overlap with LSST observing regions}
\label{subsec:rubin_overlap}

As an application of the compiled catalog, we also identify the subset of known CL-AGN that fall within Rubin/LSST observing regions. A detailed Rubin-focused analysis of known CL-AGN located within the Rubin Deep Drilling Fields was presented in our earlier work \citep{camus2026known}. Here, we include the Rubin overlap only as a practical demonstration of how the broader multiwavelength database can be used for time-domain follow-up planning. Briefly, to evaluate the Rubin accessibility of the compiled CL-AGN sample, we compared the source positions with the Rubin footprint and cadence information. We used the Rubin {\tt baseline\_v5.1.0\_10yrs} cadence simulation \citep[see e.g.,][]{2026arXiv260121769P} and selected high-cadence Wide--Fast--Deep and Deep-Drilling Field regions using a visit-count threshold of $N \geq 804$ visits.\\

Using the updated optical parent catalog, we find 70 spectroscopically confirmed CL-AGN within the LSST Wide--Fast--Deep footprint and 5 confirmed CL-AGN within the Rubin Deep Drilling Fields. The DDF sources are located in the COSMOS and XMM-LSS regions, where ongoing and future high-cadence Rubin monitoring can be combined with extensive archival multiwavelength data \citep{XMMLSS2004JCAP...09..011P,COSMOS2007ApJS..172....1S, COSMOS2010ApJ...716..348B, XMMLSS2016A&A...592A...1P, COSMOS2016ApJS..224...24L}. These objects, therefore, provide useful reference targets for studying how optical changing-look behavior connects to X-ray, infrared, ultraviolet, and radio properties.\\

The Rubin overlap should not be interpreted as a prediction of the full future Rubin CL-AGN yield. Instead, it identifies the currently known sources that Rubin will be able to monitor and provides a reference sample against which new Rubin-selected changing-look candidates can be compared. This is particularly relevant because the current confirmed CL-AGN sample is still strongly biased toward northern and equatorial surveys, while Rubin will repeatedly monitor a large area of the southern sky.\\

As new Rubin alerts, through the alert brokers\footnote{\url{https://rubinobservatory.org/for-scientists/data-products/alerts-and-brokers}}, and light curves become available, the database can be updated to include new candidates, revised classifications, and additional multiwavelength information. In this sense, the Rubin overlap is not treated here as an independent result separate from the catalog, but as one example of the catalog's utility as an expandable reference database in the Rubin era.\\

\section{Multiwavelength data and cross-matching}
\label{sec:crossmatch}

After defining the optical parent catalog and its overlap with Rubin observing regions, we extended the sample with ancillary information across the electromagnetic spectrum. The goal of this step was not to reselect CL-AGN, but to connect each known optical source with existing observations that can support follow-up planning, sample comparison, and future physical interpretation.\\

The data sets used here have very different angular resolutions,
positional uncertainties, sky coverage, and selection functions.
For this reason, we did not force a single identical matching strategy
on all wavelength regimes. Instead, we used a common probabilistic
framework where appropriate, while treating special cases separately.
For most survey catalogs with well-defined source positions, we used
the Bayesian cross-matching code \texttt{nway} \citep{salvato2018finding}.
Unless stated otherwise, we retained strong associations when the
preferred counterpart satisfied \texttt{match\_flag = 1},
\texttt{p\_any} $\geq 0.8$, and \texttt{p\_i} $\geq 0.8$. They quantify how confident the algorithm is that a candidate counterpart is the correct association, taking into account positional uncertainties, source densities, and optional priors: (i) \texttt{match\_flag = 1} is a quality flag assigned by {\tt nway} wherein the matched source is the preferred (best) counterpart among all candidates and has the highest probability. For completeness, \texttt{match\_flag = 2} reports a secondary (less likely) counterpart, and \texttt{match\_flag = 0} happens when no acceptable counterpart is identified; (ii) \texttt{p\_any} is the reliability of the $1-\texttt{p\_any}$ association. Equivalently, 1 - \texttt{p\_any}, is the probability that none of the catalog objects within the search radius is actually associated (i.e., the source is a blank field); and (iii) \texttt{p\_i} is the posterior probability for a specific candidate. Together, they ensure that (a) the counterpart is the preferred one, (b) a genuine counterpart is likely present in the comparison catalog, and (c) the preferred counterpart is itself identified with at least 80\% confidence. Here, the choice of 0.8 as a threshold is based on the earlier studies in X-ray counterpart catalogs such as eROSITA, XMM-Newton, and Chandra \citep{salvato2018finding}. The exact threshold depends on whether completeness or purity is prioritized, but 0.8 is a widely accepted compromise for producing reliable counterpart catalogs. For each layer, we stored counterpart identifiers, angular separations,
match probabilities, detection flags, and basic photometric or flux
information when available.\\

\begin{figure*}[!htb]
\centering
    \includegraphics[width=0.85\linewidth]{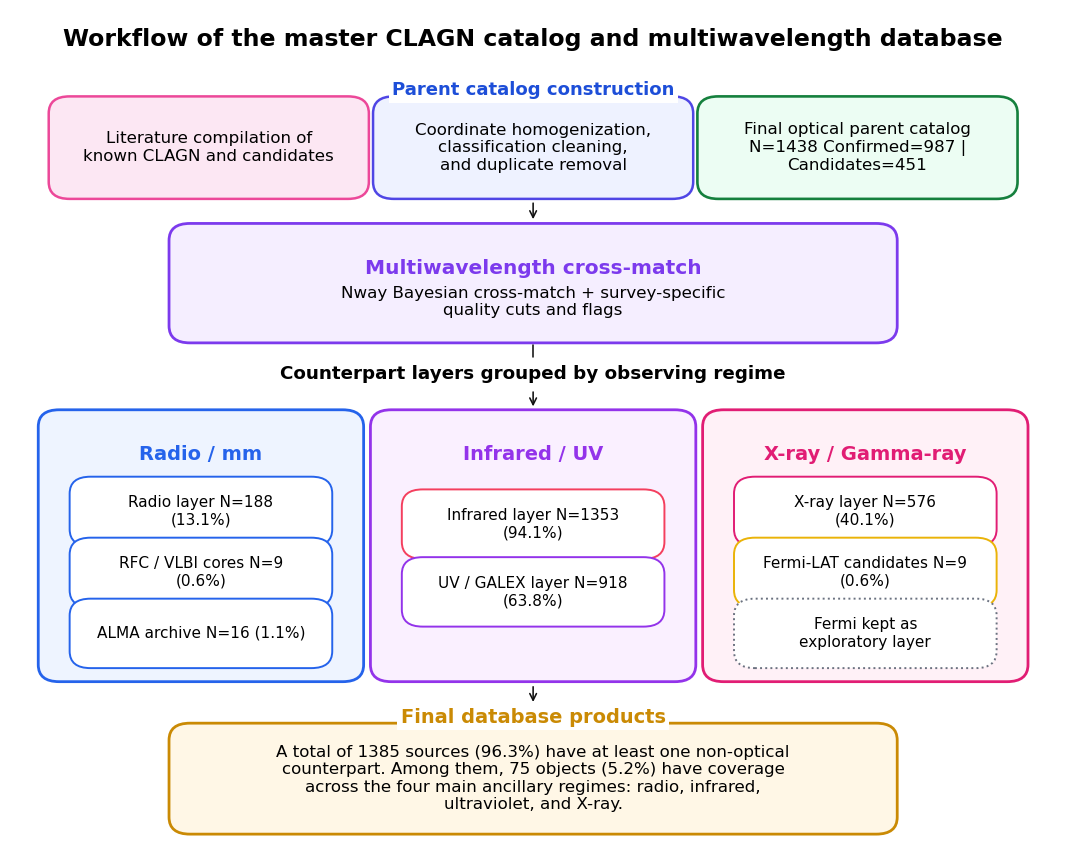}
    \caption{
Schematic overview of the construction of the master CL-AGN catalog and its multiwavelength extension. 
The upper row summarizes the compilation of literature CL-AGN samples, coordinate homogenization, classification cleaning, and duplicate removal used to define the optical parent catalog of 1,438 sources. 
The middle block shows the multiwavelength cross-match step, performed using the Bayesian cross-matching code \texttt{nway} together with survey-specific quality cuts and flags. 
The lower blocks group the ancillary layers by observing regime: radio/mm, infrared/UV, and X-ray/gamma-ray. 
The UV layer shown in the workflow corresponds to the broad GALEX coverage flag; the more conservative GALEX subset is discussed separately in the ultraviolet cross-match summary. 
Percentages are relative to the full optical parent catalog. 
Sub-layers are not mutually exclusive because a single CL-AGN can have counterparts in multiple surveys or wavelength regimes. 
A total of 1,385 sources have at least one non-optical counterpart, while 75 sources have coverage across the four main ancillary regimes: radio, infrared, ultraviolet, and X-ray. 
The Fermi-LAT associations are kept as an exploratory gamma-ray candidate layer and are not included in the main full-regime coverage statistic.
}
    \label{fig:catalog_workflow}
\end{figure*}

The full workflow of the parent-catalog construction, multiwavelength cross-matching, and final database products is summarized in Figure~\ref{fig:catalog_workflow}.\\

For traceability, the main catalog access points used for the cross-matching are listed here. 
The radio catalogs were accessed through VizieR using VLASS\footnote{\url{https://vizier.cds.unistra.fr/viz-bin/VizieR?-source=J/ApJS/255/30}}, FIRST\footnote{\url{https://vizier.cds.unistra.fr/viz-bin/VizieR?-source=VIII/92}}, NVSS\footnote{\url{https://vizier.cds.unistra.fr/viz-bin/VizieR?-source=VIII/65}}, RACS\footnote{\url{https://vizier.cds.unistra.fr/viz-bin/VizieR?-source=J/other/PASA/38.58}}, SUMSS\footnote{\url{https://vizier.cds.unistra.fr/viz-bin/VizieR?-source=VIII/81B}}, and LoTSS DR1\footnote{\url{https://vizier.cds.unistra.fr/viz-bin/VizieR?-source=J/A\%2BA/622/A1}}. 
The RFC/VLBI layer used the rfc\_2026a\footnote{\url{https://astrogeo.sciencecloud.nasa.gov/sol/rfc_2026a}} release, and the ALMA layer was queried through the ALMA Science Archive\footnote{\url{https://almascience.eso.org/aq/}}. 
The infrared catalogs were accessed through VizieR using AllWISE\footnote{\url{https://vizier.cds.unistra.fr/viz-bin/VizieR?-source=II/328}}, CatWISE2020\footnote{\url{https://vizier.cds.unistra.fr/viz-bin/VizieR?-source=II/365}}, 2MASS\footnote{\url{https://vizier.cds.unistra.fr/viz-bin/VizieR?-source=II/246}}, AKARI/IRC\footnote{\url{https://vizier.cds.unistra.fr/viz-bin/VizieR?-source=II/297}}, and AKARI/FIS\footnote{\url{https://vizier.cds.unistra.fr/viz-bin/VizieR?-source=II/298}}. 
The ultraviolet layer used GALEX/GUVcat AIS GR6+7\footnote{\url{https://vizier.cds.unistra.fr/viz-bin/VizieR?-source=II/335}}. 
The high-energy catalogs were accessed through ROSAT BSC\footnote{\url{https://heasarc.gsfc.nasa.gov/w3browse/all/rassbsc.html}}, ROSAT FSC\footnote{\url{https://heasarc.gsfc.nasa.gov/W3Browse/rosat/rassfsc.html}}, ROSAT 2RXS\footnote{\url{https://heasarc.gsfc.nasa.gov/W3Browse/rosat/rass2rxs.html}}, eRASS1\footnote{\url{https://vizier.cds.unistra.fr/viz-bin/VizieR?-source=J/A\%2BA/682/A34}}, XMM-Newton 4XMM-DR13\footnote{\url{https://vizier.cds.unistra.fr/viz-bin/VizieR?-source=IX/69}}, the Chandra Source Catalog through HEASARC\footnote{\url{https://heasarc.gsfc.nasa.gov/w3browse/chandra/csc.html}}, and Fermi-LAT 4FGL-DR3\footnote{\url{https://vizier.cds.unistra.fr/viz-bin/VizieR?-source=IX/67}}.\\

In the next sub-sections, we describe the cross-matching and the salient results, spanning from radio to gamma-ray.

\subsection{Radio surveys and VLBI compact cores}
\label{subsec:radio}

The radio layer was built using wide-area radio surveys together with a separate compact-core VLBI layer. For the survey component, we used VLASS \citep{gordon2021vlass}, FIRST \citep{helfand2015first}, NVSS \citep{condon1998nvss}, RACS \citep{hale2021racs}, SUMSS \citep{mauch2003sumss}, and LoTSS \citep{shimwell2019lotss}. These surveys differ in frequency, angular resolution, sensitivity, and sky coverage, so we retain individual survey flags as well as a combined radio-survey flag. This allows users to distinguish between sources detected in a specific radio catalog and sources with broader radio coverage in the database.\\

Using the wide-area survey flags alone, we identify 160 sources with at least one radio-survey counterpart, corresponding to 11.1\% of the full optical parent catalog. The broader radio/mm layer used in the global summary contains 188 sources, corresponding to 13.1\% of the catalog, once the additional compact-core and archival millimeter/submillimeter information is included. Although this fraction is much lower than the infrared or X-ray coverage, the radio-detected subset is physically important because it can highlight CL-AGN with possible jet activity, compact radio emission, or radio-loud behavior.\\

In addition to the lower-resolution radio surveys, we included the Radio Fundamental Catalogue as a VLBI compact-core layer. We used the rfc\_2026a release, which provides high-precision astrometry and correlated flux-density estimates for compact radio sources derived from VLBI observations \citep{petrov2025rfc}. RFC matches are rare in our catalog, with 9 sources, corresponding to 0.6\% of the full sample. We keep this layer separate from the broader radio-survey flag because an RFC match indicates compact, high-resolution radio emission rather than only a lower-resolution survey detection.\\

\begin{figure*}[!t]
    \centering

    \begin{minipage}[t]{0.49\textwidth}
        \centering

        \begin{minipage}[t][6cm][c]{\linewidth}
            \centering
            \small
            \resizebox{\linewidth}{!}{%
                \begin{tabular}{lcc}
                    \hline
                    Radio catalog & Strong matches & Detection fraction \\
                    \hline
                    VLASS & 91 & 6.33\% \\
                    FIRST & 89 & 6.19\% \\
                    NVSS & 97 & 6.75\% \\
                    RACS & 82 & 5.70\% \\
                    SUMSS & 7 & 0.49\% \\
                    LoTSS  & 32 & 2.23\% \\
                    \hline
                \hline
Any wide-area radio survey & 160 & 11.13\% \\
RFC/VLBI compact core & 9 & 0.63\% \\
ALMA archival coverage & 16 & 1.11\% \\
Broad radio/mm layer & 188 & 13.07\% \\
\hline
                \end{tabular}%
            }
        \end{minipage}

        \captionof{table}{
Radio and millimeter/submillimeter ancillary coverage used to define the broad radio/mm layer. 
Fractions are computed relative to the full optical parent catalog of 1,438 sources. 
The wide-area radio-survey row is the union of VLASS, FIRST, NVSS, RACS, SUMSS, and LoTSS, while the broad radio/mm layer additionally includes RFC/VLBI compact-core matches and ALMA archival coverage. 
}
        \label{tab:side_table}
    \end{minipage}
    \hfill
    \begin{minipage}[t]{0.49\textwidth}
        \centering

        \begin{minipage}[t][6cm][c]{\linewidth}
            \centering
            \includegraphics[
                width=\linewidth,
                height=5.8cm,
                keepaspectratio
            ]{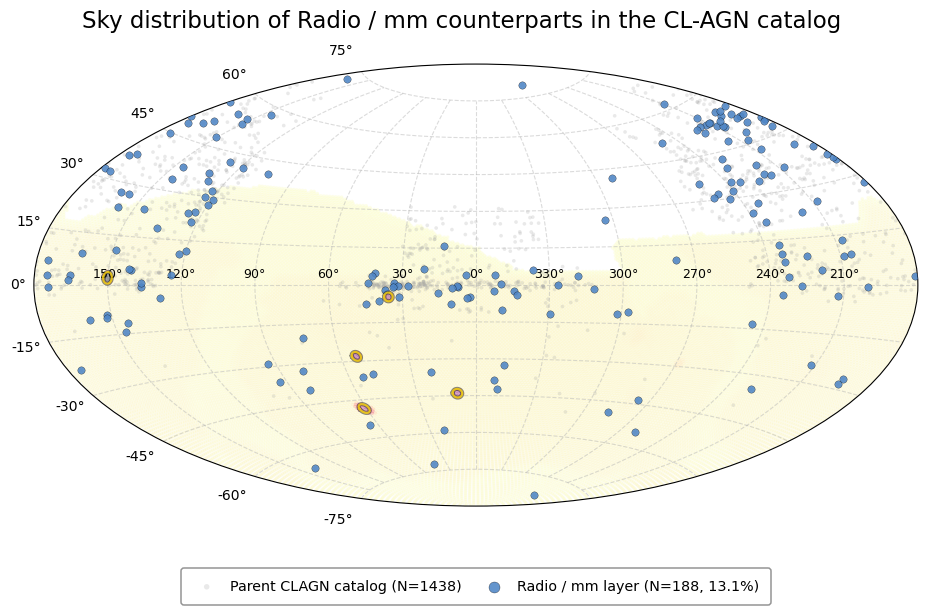}
        \end{minipage}

        \captionof{figure}{Sky distribution of the broad radio/mm ancillary layer in the master CL-AGN catalog. 
Gray points show the optical parent sample, while blue points mark the 188 sources with radio/mm ancillary coverage. 
The pale background traces the Rubin/LSST footprint, and gold open circles mark the Rubin Deep Drilling Fields. 
Catalog-level counts for the contributing radio surveys, RFC/VLBI compact-core layer, and ALMA archival layer are summarized in Table~\ref{tab:side_table}.}.

        \label{fig:side_figure}
    \end{minipage}

\end{figure*}

The radio and millimeter catalog-level statistics are summarized in Table~\ref{tab:side_table}, and the corresponding sky distribution is shown in Figure~\ref{fig:side_figure}.

\subsection{Millimeter and submillimeter archival coverage}
\label{subsec:alma}

We also searched for archival ALMA observations using the ALMA Science Archive queried through \texttt{astroquery.alma} \citep{ginsburg2019astroquery}. This layer is treated differently from the survey-based radio catalogs. An ALMA flag indicates the existence of archival observations near the optical CL-AGN position, but it does not imply that the source was observed with uniform sensitivity, frequency coverage, angular resolution, or science goals. For this reason, we treat ALMA as an archival coverage layer rather than as a homogeneous detection survey.\\

The ALMA archive query identifies 16 sources with existing observations, corresponding to 1.1\% of the full catalog. These objects form a small but useful subset for future studies of cold dust, molecular gas, or compact millimeter/submillimeter emission in CL-AGN. Because the ALMA coverage is highly non-uniform and was not designed as a complete survey of the parent sample, we keep it separate from the main radio-survey and infrared counterpart layers.

\subsection{Infrared catalogs}
\label{subsec:infrared}

The infrared layer provides the broadest multiwavelength coverage in the catalog. 
We used AllWISE \citep{cutri2013allwise} and CatWISE2020 \citep{marocco2021catwise} as the main mid-infrared data sets, 2MASS \citep{cutri2003twomass} for near-infrared photometry, and AKARI/IRC and AKARI/FIS to extend the catalog toward mid- and far-infrared wavelengths for a smaller subset of sources \citep{ishihara2010akari,yamamura2010akari}. 
These catalogs provide complementary infrared information, ranging from near-infrared stellar and host-galaxy emission to mid-infrared dust emission and, for a small number of sources, far-infrared archival coverage.\\

AllWISE and CatWISE2020 provide the main WISE-based coverage, including W1 and W2 photometry and longer-wavelength WISE measurements when available. 
We also computed WISE colors, including W1--W2, because this color is commonly used as a mid-infrared AGN diagnostic. 
2MASS adds J, H, and Ks photometry, while AKARI provides additional longer-wavelength information through the IRC mid-infrared and FIS far-infrared all-sky catalogs.\\

The final infrared layer contains 1,353 sources detected in at least one infrared catalog, corresponding to 94.1\% of the full optical sample. 
This high fraction is mainly driven by the wide sky coverage of WISE and CatWISE2020. 
Using the W1--W2 $>$ 0.8 criterion in either AllWISE or CatWISE2020, 685 sources satisfy an AGN-like mid-infrared color selection.\\

\begin{figure*}[!t]
    \centering

    \begin{minipage}[t]{0.49\textwidth}
        \centering

        \begin{minipage}[t][6cm][c]{\linewidth}
            \centering
            \small
            \resizebox{\linewidth}{!}{%
                \begin{tabular}{lcc}
                    \hline
                    Infrared layer or subset
                    & Number of sources
                    & Fraction of full catalog \\
                    \hline
                    AllWISE & 1183 & 82.27\% \\
                    CatWISE2020 & 1232 & 85.67\% \\
                    2MASS PSC & 489 & 34.01\% \\
                    AKARI/IRC & 16 & 1.11\% \\
                    AKARI/FIS & 19 & 1.32\% \\
                    Any AKARI layer & 25 & 1.74\% \\
                    \hline
                    Any infrared catalog before AKARI
                    & 1349 & 93.81\% \\
                    Any infrared catalog including AKARI
                    & 1353 & 94.09\% \\
                    New infrared sources added only by AKARI
                    & 4 & 0.28\% \\
                    \hline
                    AllWISE $W1-W2 > 0.8$
                    & 599 & 41.66\% \\
                    CatWISE2020 $W1-W2 > 0.8$
                    & 556 & 38.66\% \\
                    Either WISE catalog, $W1-W2 > 0.8$
                    & 685 & 47.64\% \\
                    \hline
                \end{tabular}%
            }
        \end{minipage}

        \captionof{table}{Infrared ancillary coverage in the master CL-AGN catalog. 
Fractions are computed relative to the full optical parent sample of 1,438 sources. 
The table separates the individual near- and mid-infrared catalogs from the combined infrared flags, and also reports the subset satisfying the commonly used WISE color criterion $W1-W2>0.8$. 
Rows are not mutually exclusive because a single source may be detected in more than one infrared catalog.
}
        \label{tab:infrared_crossmatch_summary}
    \end{minipage}
    \hfill
    \begin{minipage}[t]{0.49\textwidth}
        \centering

        \begin{minipage}[t][6cm][c]{\linewidth}
            \centering
            \includegraphics[
                width=\linewidth,
                height=5.8cm,
                keepaspectratio
            ]{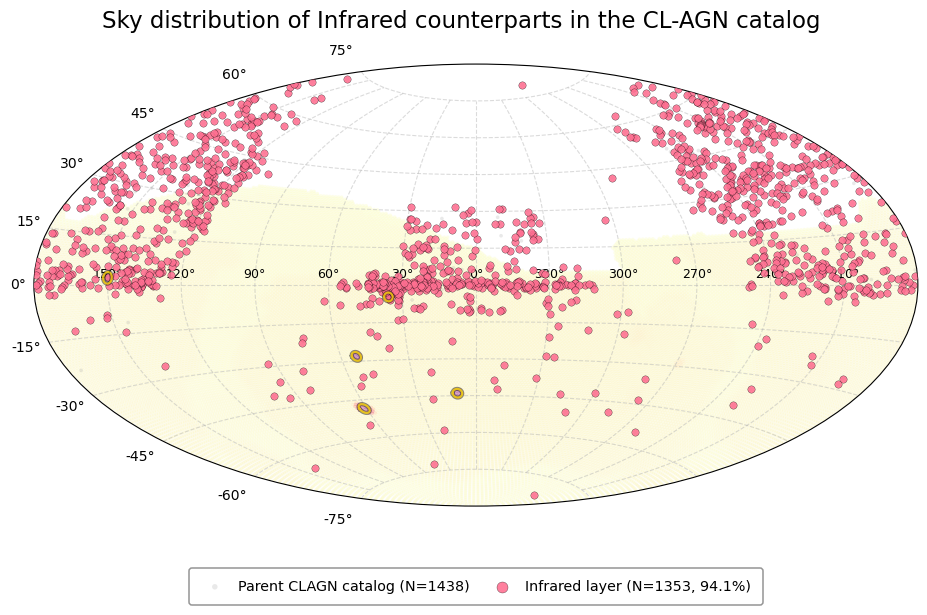}
        \end{minipage}

        \captionof{figure}{Sky distribution of the infrared ancillary layer in the master CL-AGN catalog. 
Gray points show the optical parent sample, while pink points mark sources with at least one infrared counterpart. 
The pale background traces the Rubin/LSST footprint, and gold open circles mark the Rubin Deep Drilling Fields. 
The corresponding catalog-level infrared statistics, including the WISE color-selected subset, are summarized in Table~\ref{tab:infrared_crossmatch_summary}.
}

        \label{fig:infrared_crossmatches}
    \end{minipage}

\end{figure*}

The infrared catalog-level statistics are summarized in Table~\ref{tab:infrared_crossmatch_summary}, and the corresponding sky distribution is shown in Figure~\ref{fig:infrared_crossmatches}.

\subsection{Ultraviolet data}
\label{subsec:uv}

The ultraviolet layer was constructed using the GALEX/GUVcat AIS GR6+7 catalog \citep{martin2005galex,bianchi2017guvcat}, queried through VizieR with a 5 arcsec search radius around each optical CL-AGN position. GALEX provides ultraviolet photometry in the near-ultraviolet (NUV) and far-ultraviolet (FUV) bands, offering useful information on the blue/UV emission associated with the accretion flow and recent variability in AGN.\\

GALEX required a slightly different treatment from the infrared catalogs because many plausible UV associations have good angular separations but do not always pass the same conservative \texttt{p\_any} threshold used for denser catalogs. For this reason, we keep two GALEX quality levels in the final catalog. The broad GALEX best-match layer is intended as a UV coverage indicator, while the conservative GALEX strong-match subset provides a stricter set of UV associations for analyses that require higher counterpart confidence.\\

The broad GALEX best-match layer includes 918 sources, corresponding to 63.8\% of the catalog. The conservative GALEX strong-match subset includes 139 sources, corresponding to 9.7\%. We therefore use the broad GALEX layer when discussing overall UV coverage, while retaining the strong subset for analyses that require stricter counterpart confidence. NUV photometry is available for 918 sources, while FUV photometry and FUV--NUV color are available for 655 sources. These quantities are stored in the catalog together with the GALEX match quality flag.\\

\begin{figure*}[!t]
    \centering

    \begin{minipage}[t]{0.49\textwidth}
        \centering

        \begin{minipage}[t][6cm][c]{\linewidth}
            \centering
            \small
            \resizebox{\linewidth}{!}{%
                \begin{tabular}{lcc}
                    \hline
                    Ultraviolet layer or subset
                    & Number of sources
                    & Fraction of full catalog \\
                    \hline
                    GALEX best match & 918 & 63.84\% \\
                    GALEX strong match & 139 & 9.67\% \\
                    GALEX best match only & 779 & 54.17\% \\
                    No GALEX best match & 520 & 36.16\% \\
                    Best matches with separation $>3\arcsec$
                    & 39 & 2.71\% \\
                    \hline
                    NUV photometry available & 918 & 63.84\% \\
                    FUV photometry available & 655 & 45.55\% \\
                    FUV--NUV color available & 655 & 45.55\% \\
                    \hline
                \end{tabular}%
            }
        \end{minipage}

        \captionof{table}{GALEX ultraviolet ancillary coverage in the master CL-AGN catalog. 
Fractions are computed relative to the full optical parent sample of 1,438 sources. 
The GALEX best-match layer is used as a broad UV coverage flag, while the strong-match subset provides a more conservative set of associations for analyses requiring higher counterpart confidence. 
The table also reports the availability of NUV, FUV, and FUV--NUV photometric information in the final catalog.
}

        \label{tab:uv_crossmatch_summary}
    \end{minipage}
    \hfill
    \begin{minipage}[t]{0.49\textwidth}
        \centering

        \begin{minipage}[t][6cm][c]{\linewidth}
            \centering
            \includegraphics[
                width=\linewidth,
                height=5.8cm,
                keepaspectratio
            ]{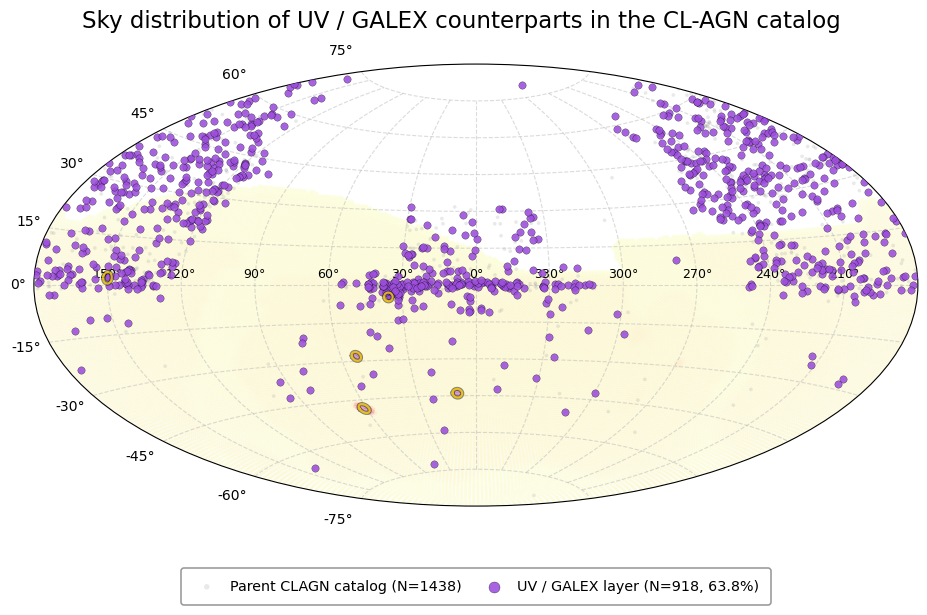}
        \end{minipage}

        \captionof{figure}{Sky distribution of the GALEX ultraviolet ancillary layer in the master CL-AGN catalog. 
Gray points show the optical parent sample, while purple points mark sources included in the broad GALEX best-match layer. 
The pale background traces the Rubin/LSST footprint, and gold open circles mark the Rubin Deep Drilling Fields. 
The GALEX quality subsets and UV photometric availability are summarized in Table~\ref{tab:uv_crossmatch_summary}.
}
        \label{fig:galex_crossmatch}
    \end{minipage}

\end{figure*}

The GALEX match-quality subsets and ultraviolet photometric availability are summarized in Table~\ref{tab:uv_crossmatch_summary}, and the corresponding sky distribution is shown in Figure~\ref{fig:galex_crossmatch}.

\subsection{X-ray catalogs}
\label{subsec:xray}

The X-ray layer combines information from ROSAT, eROSITA/eRASS1, XMM-Newton, and Chandra. 
For ROSAT, we used three all-sky catalog products: the ROSAT Bright Source Catalogue \citep{voges1999rosat}, the ROSAT Faint Source Catalogue \citep{voges2000rosat}, and the second ROSAT all-sky survey source catalog, 2RXS \citep{boller2016rosat}. 
For eROSITA, we used the first SRG/eROSITA all-sky survey catalogs, eRASS1 \citep{merloni2024erass1}. 
These wide-area X-ray surveys were complemented with pointed X-ray source catalogs from XMM-Newton and Chandra.\\

For XMM-Newton, we used the 4XMM-DR13 source catalog \citep{webb2020xmm}, queried through VizieR. 
For Chandra, we used the Chandra Source Catalog queried through HEASARC \citep{evans2010csc}. 
Because the angular resolution, depth, and sky coverage vary substantially between these facilities, we preserve individual catalog flags rather than collapsing all X-ray information into a single quantity at the catalog-query stage. 
We then define a combined X-ray flag to identify sources detected in at least one X-ray layer.\\

The final combined X-ray layer contains 576 sources with at least one X-ray counterpart, corresponding to 40.1\% of the full optical catalog. 
This makes X-rays one of the most important ancillary wavelength regimes in the database, after the infrared layer.\\

\begin{figure*}[!t]
    \centering

    \begin{minipage}[t]{0.49\textwidth}
        \centering

        \begin{minipage}[t][6cm][c]{\linewidth}
            \centering
            \small
            \resizebox{\linewidth}{!}{%
                \begin{tabular}{lcc}
                    \hline
                    X-ray layer or subset
                    & Number of sources
                    & Fraction of full catalog \\
                    \hline
                    ROSAT BSC & 97 & 6.75\% \\
                    ROSAT FSC & 156 & 10.85\% \\
                    ROSAT 2RXS & 256 & 17.80\% \\
                    Any ROSAT layer & 285 & 19.82\% \\
                    \hline
                    eRASS1 main catalog & 266 & 18.50\% \\
                    eRASS1 supplementary catalog & 10 & 0.70\% \\
                    eRASS1 hard catalog & 32 & 2.23\% \\
                    Any eRASS1 layer & 276 & 19.19\% \\
                    \hline
                    XMM-Newton 4XMM-DR13 & 135 & 9.39\% \\
                    Chandra CSC & 89 & 6.19\% \\
                    \hline
                    Any X-ray layer & 576 & 40.06\% \\
                    \hline
                \end{tabular}%
            }
        \end{minipage}

        \captionof{table}{X-ray ancillary coverage in the master CL-AGN catalog. 
Fractions are computed relative to the full optical parent sample of 1,438 sources. 
The table separates the ROSAT, eRASS1, XMM-Newton, and Chandra contributions from the combined X-ray flag. 
Rows are not mutually exclusive because a single source may have counterparts in more than one X-ray catalog.}
        \label{tab:xray_crossmatch_summary}
    \end{minipage}
    \hfill
    \begin{minipage}[t]{0.49\textwidth}
        \centering

        \begin{minipage}[t][6cm][c]{\linewidth}
            \centering
            \includegraphics[
                width=\linewidth,
                height=5.8cm,
                keepaspectratio
            ]{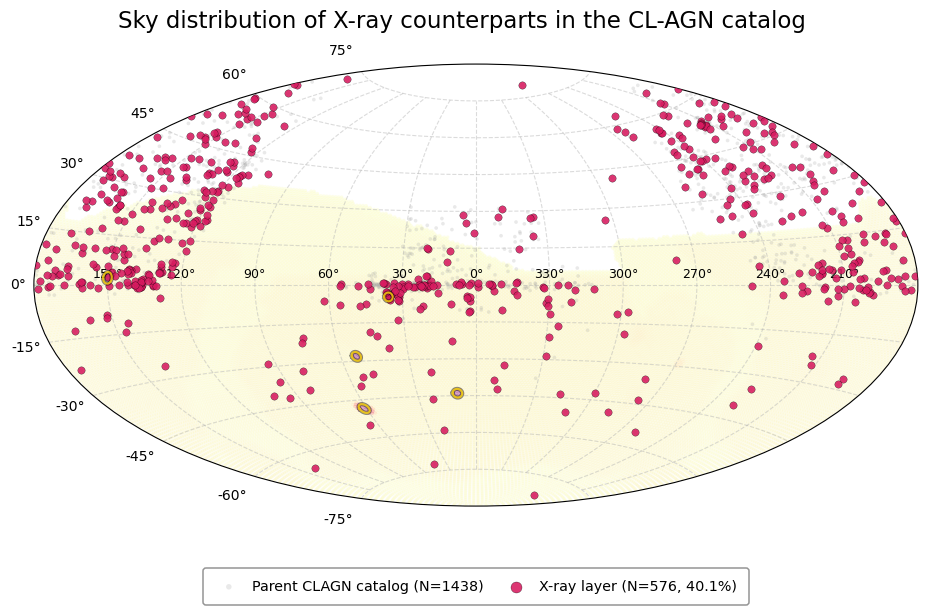}
        \end{minipage}

        \captionof{figure}{Sky distribution of the combined X-ray ancillary layer in the master CL-AGN catalog. 
Gray points show the optical parent sample, while magenta points mark sources with at least one X-ray counterpart. 
The pale background traces the Rubin/LSST footprint, and gold open circles mark the Rubin Deep Drilling Fields. 
The catalog-level X-ray statistics are summarized in Table~\ref{tab:xray_crossmatch_summary}.
}
        \label{fig:xray_crossmatches}
    \end{minipage}

\end{figure*}

The X-ray catalog-level statistics are summarized in Table~\ref{tab:xray_crossmatch_summary}, and the corresponding sky distribution is shown in Figure~\ref{fig:xray_crossmatches}.

\subsection{Exploratory gamma-ray candidate associations}
\label{subsec:gamma}

\begin{figure*}[!t]
    \centering

    \begin{minipage}[t]{0.49\textwidth}
        \centering

        \begin{minipage}[t][6cm][c]{\linewidth}
            \centering
            \small
            \resizebox{\linewidth}{!}{%
                \begin{tabular}{lcc}
                    \hline
                    Fermi-LAT layer or subset
                    & Number of sources
                    & Fraction of full catalog \\
                    \hline
                    4FGL-DR3 candidate rows within $30'$
                    & 189 & -- \\
                    Optical sources with a 4FGL candidate within $30'$
                    & 180 & 12.5\% \\
                    Best 4FGL candidate per optical source
                    & 180 & 12.5\% \\
                    Candidate associations inside the 95\% error ellipse
                    & 9 & 0.6\% \\
                    \hline
                    AGN/blazar-like gamma-ray candidates
                    & 5 & 0.3\% \\
                    Unassociated or unknown gamma-ray candidates
                    & 3 & 0.2\% \\
                    Likely non-AGN gamma-ray association
                    & 1 & 0.1\% \\
                    \hline
                    Fermi candidates with radio-survey support
                    & 2 & 0.1\% \\
                    Fermi candidates with RFC/VLBI compact-core match
                    & 0 & 0.0\% \\
                    \hline
                \end{tabular}%
            }
        \end{minipage}

        \captionof{table}{Exploratory Fermi-LAT 4FGL-DR3 gamma-ray candidate-association layer. 
Fractions are computed relative to the full optical parent sample of 1,438 sources, except for the raw 4FGL-DR3 candidate rows, which are not unique optical sources. 
The final candidate layer includes only optical CL-AGN positions that fall within the corresponding 95\% Fermi-LAT error ellipse after keeping the best 4FGL-DR3 candidate per optical source. 
These associations are retained for source-by-source inspection and are not included in the main secure multiwavelength counterpart statistics.
        }
        \label{tab:fermi_crossmatch_summary}
    \end{minipage}
    \hfill
    \begin{minipage}[t]{0.49\textwidth}
        \centering

        \begin{minipage}[t][6cm][c]{\linewidth}
            \centering
            \includegraphics[
                width=\linewidth,
                height=5.8cm,
                keepaspectratio
            ]{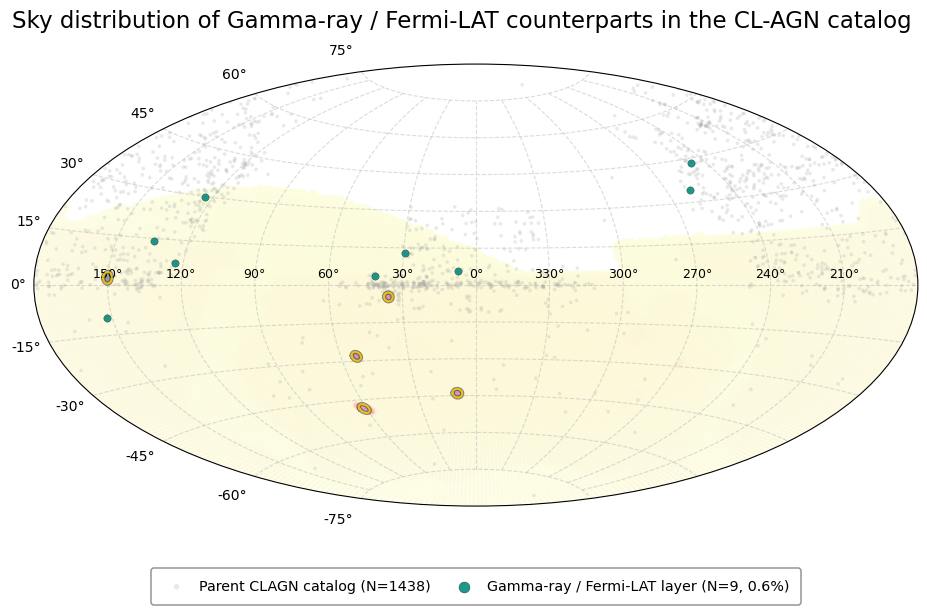}
        \end{minipage}

        \captionof{figure}{
Sky distribution of the exploratory Fermi-LAT candidate-association layer in the master CL-AGN catalog. 
Gray points show the optical parent sample, while teal points mark the 9 sources whose best 4FGL-DR3 candidate lies within the corresponding 95\% Fermi-LAT error ellipse. 
The pale background traces the Rubin/LSST footprint, and gold open circles mark the Rubin Deep Drilling Fields. 
The Fermi-LAT candidate-association statistics are summarized in Table~\ref{tab:fermi_crossmatch_summary}.
}
\label{fig:fermi_crossmatches}
\end{minipage}

\end{figure*}

Finally, we tested an exploratory gamma-ray layer using the Fermi-LAT 4FGL-DR3 catalog \citep{abdollahi2022fermi}. 
This layer was treated with particular caution because Fermi-LAT localization regions are much larger than the positional uncertainties of the optical, infrared, ultraviolet, X-ray, and radio catalogs. 
For this reason, we do not refer to these matches as secure counterparts. 
Instead, we keep them as candidate gamma-ray associations.\\

For each optical CL-AGN, we searched for 4FGL-DR3 sources within 30 arcmin. We then required the optical position to fall inside the 95\% Fermi-LAT error ellipse, using the semi-major axis, semi-minor axis, and position angle reported in the 4FGL catalog. This initial search returned 189 candidate rows around 180 optical sources. After applying the 95\% error-ellipse requirement and keeping the best candidate per optical source, 9 candidate associations remain. Of these, 5 have AGN or blazar-like Fermi classes, 3 are unassociated or unknown, and 1 is classified as a millisecond pulsar.\\

We keep this layer separate from the main multiwavelength counterpart layers because the associations require manual validation. None of the 9 Fermi candidates overlap with the RFC/VLBI compact-core layer, and only 2 are supported by the radio-survey layers used in this work. The millisecond pulsar case is flagged as a likely non-AGN gamma-ray association and requires a second inspection.\\

Because the Fermi-LAT layer is exploratory and based on comparatively large localization regions, these candidate associations are not included in the main secure multiwavelength counterpart statistics. Instead, they are retained as flags for future source-by-source inspection.\\

The exploratory Fermi-LAT candidate-association statistics are summarized in Table~\ref{tab:fermi_crossmatch_summary}, and the corresponding sky distribution is shown in Figure~\ref{fig:fermi_crossmatches}.

\section{Global multiwavelength coverage}
\label{sec:mw_results}

After constructing the individual wavelength layers, we combined them into a set of global multiwavelength flags. These flags allow us to identify which CL-AGN have ancillary information in each broad wavelength domain, independent of the specific survey in which the counterpart was found. In this section, we summarize the overall coverage of the catalog and identify the subsets with the richest multiwavelength information.\\

A compact summary of the ancillary layers included in the database is given in Table~\ref{tab:multiwavelength_simple_summary}.\\

The strongest coverage comes from the infrared. In total, 1,353 sources have a counterpart in at least one infrared catalog, corresponding to 94.1\% of the full optical catalog. This high fraction is mainly driven by the wide-area coverage of AllWISE and CatWISE2020. The ultraviolet layer also provides substantial coverage when using the broad GALEX best-match sample, with 918 sources, or 63.8\% of the catalog. The conservative GALEX strong-match subset is smaller, with 139 sources, or 9.7\%, and is retained for stricter analyses.\\

The combined X-ray layer identifies 576 sources with at least one X-ray counterpart, corresponding to 40.1\% of the catalog. This makes X-rays one of the most important ancillary regimes after the infrared and broad UV layers. In contrast, radio detections are less common: 160 sources are detected in at least one wide-area radio survey, corresponding to 11.1\% of the sample. The broader radio/mm layer contains 188 sources, or 13.1\%, after including the RFC/VLBI compact-core and ALMA archival coverage layers. Although the radio fraction is lower, this subset is physically important because it selects sources with possible jet activity, compact radio emission, or radio-loud behavior.\\

We also include two smaller but useful ancillary layers. The RFC/VLBI compact-core layer contains 9 sources, corresponding to 0.6\% of the catalog, while the ALMA archival coverage layer contains 16 sources, corresponding to 1.1\%. These are not large statistical subsamples, but they are valuable for identifying objects with high-resolution radio information or existing millimeter/submillimeter observations.\\

Overall, most sources in the catalog have at least one non-optical counterpart. When using the broad GALEX best-match layer, 1,385 sources have at least one counterpart outside the optical, corresponding to 96.3\% of the catalog. Even when using only the conservative GALEX strong-match subset, 1,368 sources still have at least one non-optical counterpart, corresponding to 95.1\%. This shows that the catalog is not only a list of optical CL-AGN, but also a practical multiwavelength resource for selecting sources with existing ancillary data.\\

A particularly useful subset is formed by sources with simultaneous coverage in radio, X-ray, infrared, and ultraviolet data. Using the broad GALEX layer, we find 75 sources with coverage in all four domains. Using the conservative GALEX strong-match subset, this number decreases to 13 sources. The 75-source broad-UV subset is especially useful for follow-up planning and future population studies, because these objects already have multiwavelength information tracing different physical components of the AGN system.\\

The exploratory Fermi-LAT layer is not included in the main non-optical counterpart statistics because the associations are less secure than the lower-energy matches. As described in Section~\ref{subsec:gamma}, we identify 9 candidate gamma-ray associations inside the 95\% Fermi-LAT error ellipse, including 5 AGN/blazar-like candidates. These sources are kept as a separate high-energy candidate-association layer and should be inspected individually before being used as physical gamma-ray counterparts.\\


\section{The CL-AGN database}
\label{sec:database}

The compiled catalog is intended not only as the data product accompanying this paper, but also as the foundation of a public, continuously expandable CL-AGN database. For each source, the database provides the essential information needed to identify the object, trace its discovery and literature references, and assess its available multiwavelength coverage.\\

Each entry includes the source identifier, primary name, coordinates, redshift (when available), optical classification, confirmed/candidate status, and relevant literature references. It further incorporates Rubin/LSST footprint flags together with multiwavelength information spanning the radio, VLBI, millimeter/submillimeter, infrared, ultraviolet, X-ray, and exploratory gamma-ray regimes described above. Whenever available, the database records counterpart identifiers, angular separations, Bayesian match probabilities, photometric or flux measurements, and associated quality flags.\\

The database is designed to support flexible queries, allowing users to search by source name or coordinates, select confirmed CL-AGN, identify objects within the Rubin footprint, or retrieve samples based on available ancillary data (e.g., X-ray detections, radio counterparts, GALEX coverage, or combined radio--mm--IR--UV--X-ray--$\gamma$-ray observations).\\

Motivated by the anticipated rapid growth of the CL-AGN population in the Rubin era, the database is designed for continual updates as new sources are discovered, classifications are revised, and additional multiwavelength observations become available.\\

\begin{figure*}[!htb]
    \centering
    \includegraphics[width=\linewidth]{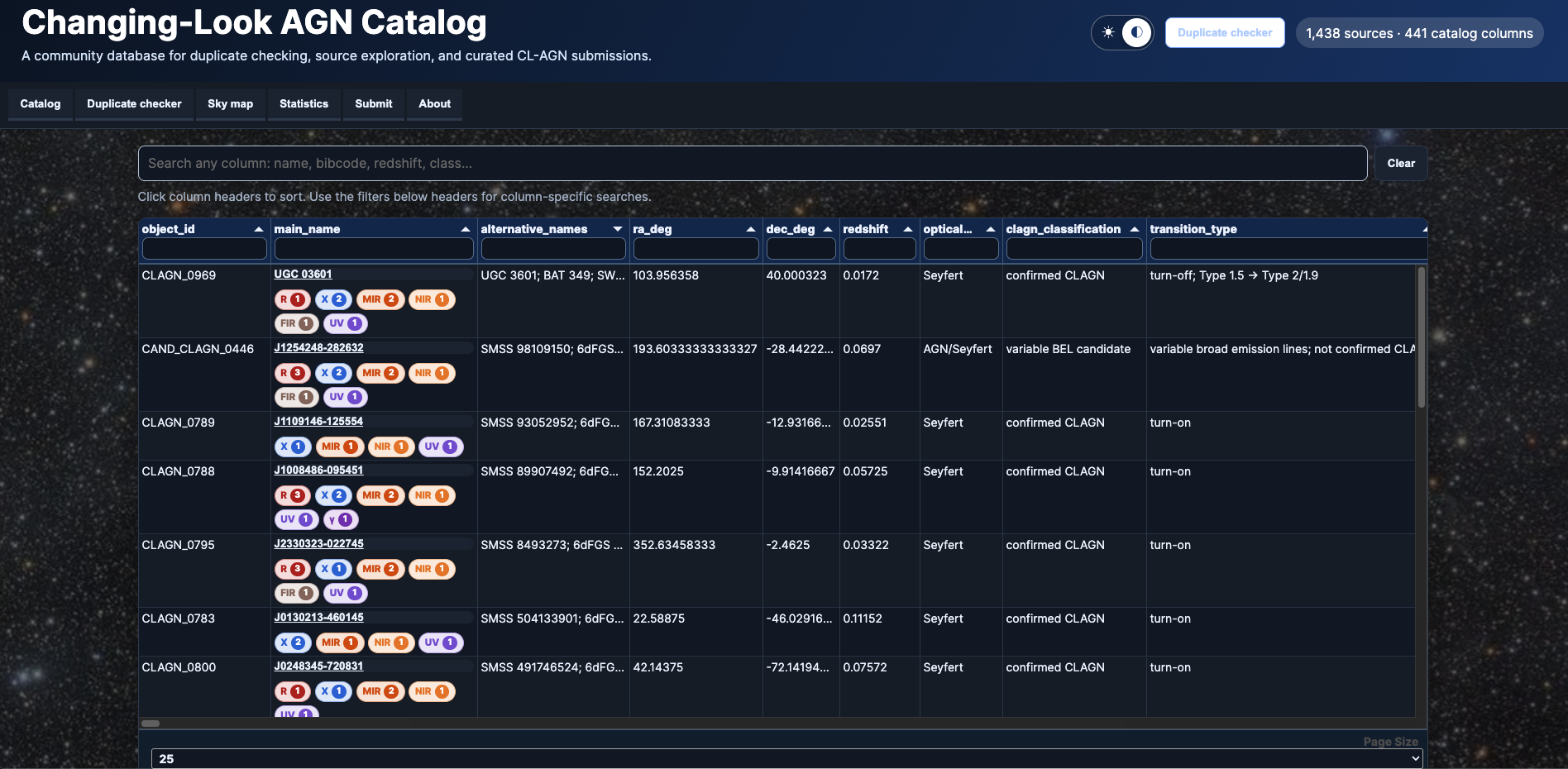}
    \caption{Main interface of the \href{https://swayamtrupta.github.io/clagn-catalog-public/}{public Changing-look AGN (CL-AGN) database}, showing the interactive catalog browser with searchable and filterable source properties, multiwavelength coverage indicators, and tools for duplicate checking, sky visualization, statistics, and community submissions.}
    \label{fig:main_page}
\end{figure*}

To maximize its scientific utility, we have also developed a dedicated public web interface\footnote{\url{https://swayamtrupta.github.io/clagn-catalog-public/}} that enables interactive exploration of the catalog while facilitating community contributions. The interface provides rapid access to individual sources, visualization of their observational properties, and direct links to publicly available data products, serving as an evolving community resource rather than a static catalog (see Figure \ref{fig:main_page}).\\


\subsection{Interactive Source Pages}

\begin{figure}
    \centering
    \includegraphics[width=\linewidth]{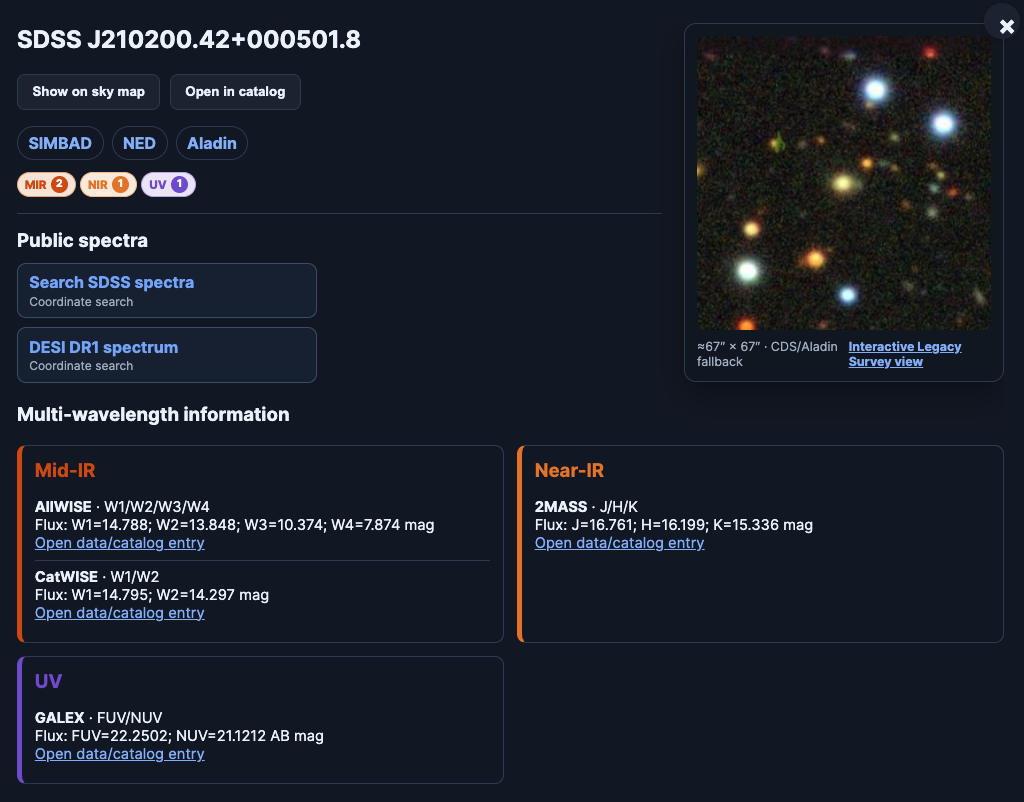}
    \caption{Example of the source-specific information panel in the public CL-AGN database. The pop-up provides a consolidated view of an individual CL-AGN, including external database links (e.g., SIMBAD, NED, and Aladin), public spectroscopic resources (e.g., SDSS and DESI), an interactive sky image, and available multiwavelength data products with direct links to the corresponding public catalogs.}
    \label{fig:pop-up}
\end{figure}

Each catalog entry can be opened through an interactive pop-up window that consolidates the available information for an individual object into a single view (see Figure \ref{fig:pop-up}). The source page includes an interactive optical image viewer centered on the target, allowing users to inspect the surrounding field and access external sky-survey visualizations. The default image viewer - the DESI Legacy Imaging Survey \citep{Dey2019AJ....157..168D} is queried first. We incorporate multiple fallback options (CDS/SIMBAD and NED) that submit image requests via these ancillary channels when the primary server is offline. In addition, publicly available spectra (e.g., SDSS, DESI DR1, 6dF Galaxy Survey, ESO archive, etc.) associated with the source are linked directly from the interface whenever available, enabling rapid inspection of the spectroscopic observations without requiring independent archive searches. The multi-wavelength properties of each source are presented through compact, color-coded tags that summarize data availability across the radio, infrared, optical, ultraviolet, and X-ray regimes. Selecting these tags expands the corresponding measurements and survey associations, providing an organized summary of the available observational coverage.

\subsection{Catalog Exploration and Visualization}

\begin{figure}
    \centering
    \includegraphics[width=\linewidth]{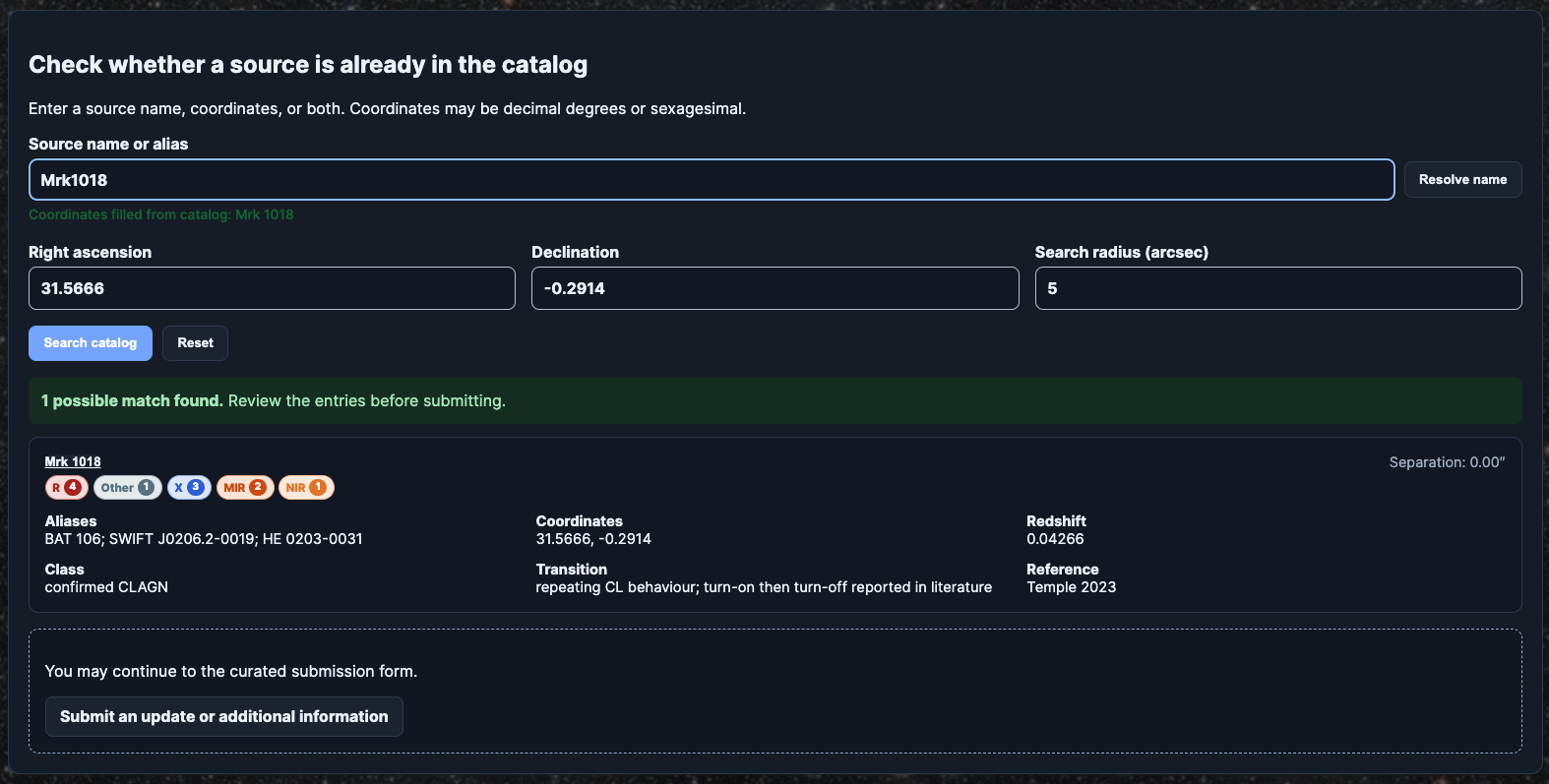}
    \caption{Duplicate-checking interface of the public CL-AGN database. The tool enables users to search the catalog by source name or coordinates within a user-defined angular radius to identify potential duplicate entries before submission. Matching sources are summarized with their classification, aliases, redshift, transition type, literature reference, and multiwavelength coverage, while providing a direct pathway for submitting updates or additional information to existing catalog entries.}
    \label{fig:duplicate-checker}
\end{figure}

Beyond individual source inspection, the website provides several interactive tools for exploring the catalog as a whole. A duplicate-checker utility allows users to search the existing database before submitting new candidates, reducing duplicate entries and maintaining catalog consistency (see Figure \ref{fig:duplicate-checker}). Interactive sky maps enable visualization of the spatial distribution of confirmed and candidate CL-AGNs together with their survey footprints, while dedicated statistics pages summarize the current properties of the catalog through dynamically generated demographic plots and survey statistics. These visualization tools provide users with an immediate overview of the catalog content and its multi-wavelength observational completeness.

\subsection{Community Contributions}

The CL-AGN catalog is designed as a community-driven resource that can evolve alongside the rapidly growing literature. The website provides a submission portal through which users can report newly identified changing-look AGNs or propose updates to existing entries, including revised classifications, additional spectroscopic observations, or newly available multi-wavelength data products. Submitted information is reviewed before incorporation into the public catalog to ensure consistency and data quality. This framework enables the catalog to remain continuously updated while maintaining a curated and homogeneous database suitable for statistical studies and follow-up observations.\\

 We envision that the AGN community at large will benefit from this database and will help to further this initiative and help maintain the ever-growing numbers of detected and confirmed CL-AGNs.\\

\section{Database use, limitations, and future updates}
\label{sec:discussion}

The catalog developed in this work provides a practical framework for exploring the known population of changing-look AGN. In addition to the basic source information, it preserves the literature provenance, confirmation status, Rubin/LSST footprint membership, and availability of ancillary observations at different wavelengths. These fields allow the sample to be filtered according to a particular scientific goal, for example by selecting spectroscopically confirmed CL-AGN, candidate systems, sources observable by Rubin, or objects with radio, infrared, ultraviolet, X-ray, or millimeter coverage.\\ 

\subsection{Catalog Reliability, Cross-Matching Strategy, and Limitations}

An important aspect of the catalog construction is that the reliability of a counterpart cannot be evaluated using the same positional criterion at every wavelength. The surveys included here differ substantially in angular resolution, source density, sensitivity, and positional accuracy. Consequently, a fixed-radius nearest-neighbor match would produce associations with different levels of reliability across the various catalog layers. We therefore used the probabilistic cross-matching code \texttt{nway} for the principal survey-based associations and applied survey-dependent probability and quality criteria. Search radii of 2, 5, and 10 arcsec were also examined where appropriate. These tests were used to identify the point at which increasing the search radius mainly introduced ambiguous or potentially spurious counterparts, and they informed the final definition of the survey-specific flags.\\

Despite these precautions, several limitations should be considered when using the database. The optical parent sample is compiled from heterogeneous literature studies that adopted different selection methods, observing strategies, and classification criteria. Its completeness is therefore determined by the sources reported to date and should not be interpreted as a complete census of the intrinsic CL-AGN population. Similarly, the absence of a multiwavelength counterpart does not necessarily indicate that a source is intrinsically faint at that wavelength. It may instead reflect incomplete sky coverage, insufficient survey depth, source variability, or the angular resolution of the corresponding catalog.\\

The counterpart flags should also be interpreted in the context of their wavelength-dependent uncertainties. Probabilistic associations provide a consistent way to rank possible counterparts, but they do not replace source-by-source validation for detailed physical studies. This is particularly relevant for catalogs with relatively broad localization regions. For this reason, the Fermi-LAT results are presented as an exploratory candidate-association layer rather than being included directly in the main counterpart statistics. Follow-up observations or additional supporting evidence would be required to establish secure gamma-ray associations.\\

The database is expected to evolve as new CL-AGN are discovered and existing classifications are revised. Future releases will incorporate newly reported sources, updated coordinates and redshifts, additional references, and improved ancillary information. Extending the catalog to larger candidate samples, including samples from surveys such as CLAS+ and upcoming data releases from SDSS-V and DESI, will also help broaden its coverage, particularly in the southern sky. These additions will become increasingly relevant as Rubin/LSST begins to confirm existing candidates and identify large numbers of variable AGN candidates.\\

More detailed applications, including population-level comparisons, light-curve modeling, spectroscopic follow-up, and machine-learning searches, are beyond the scope of the present work. Nevertheless, these are good avenues worth exploring in future investigations.\\ 

\subsection{ugrizy Photometry of a CL-AGN with LSSTCam: A Rubin-Only Case Study}
\label{sec:rubin_source}

\begin{figure*}[!t]
    \centering
    \includegraphics[width=\linewidth]{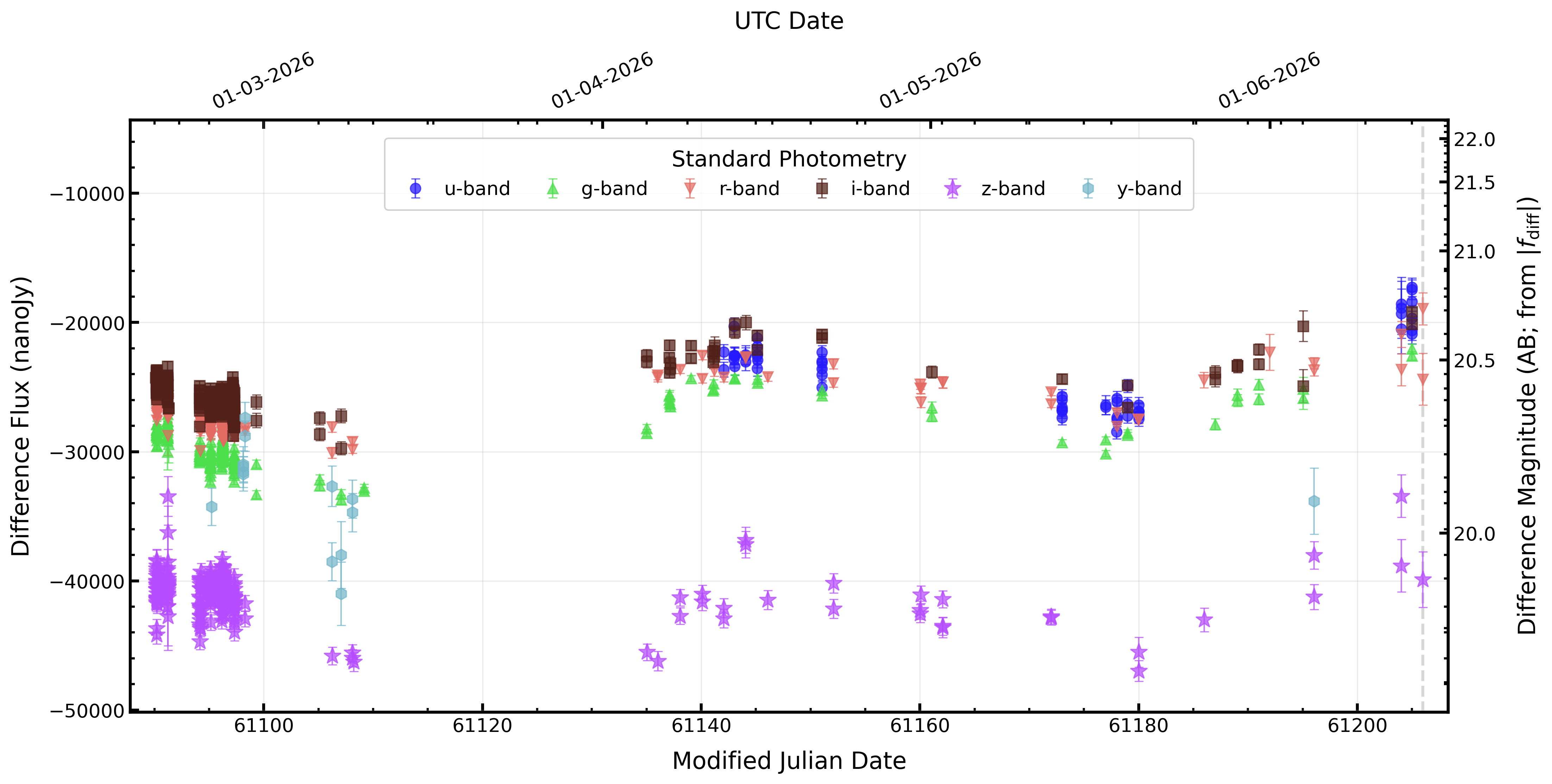}
    \includegraphics[width=\linewidth]{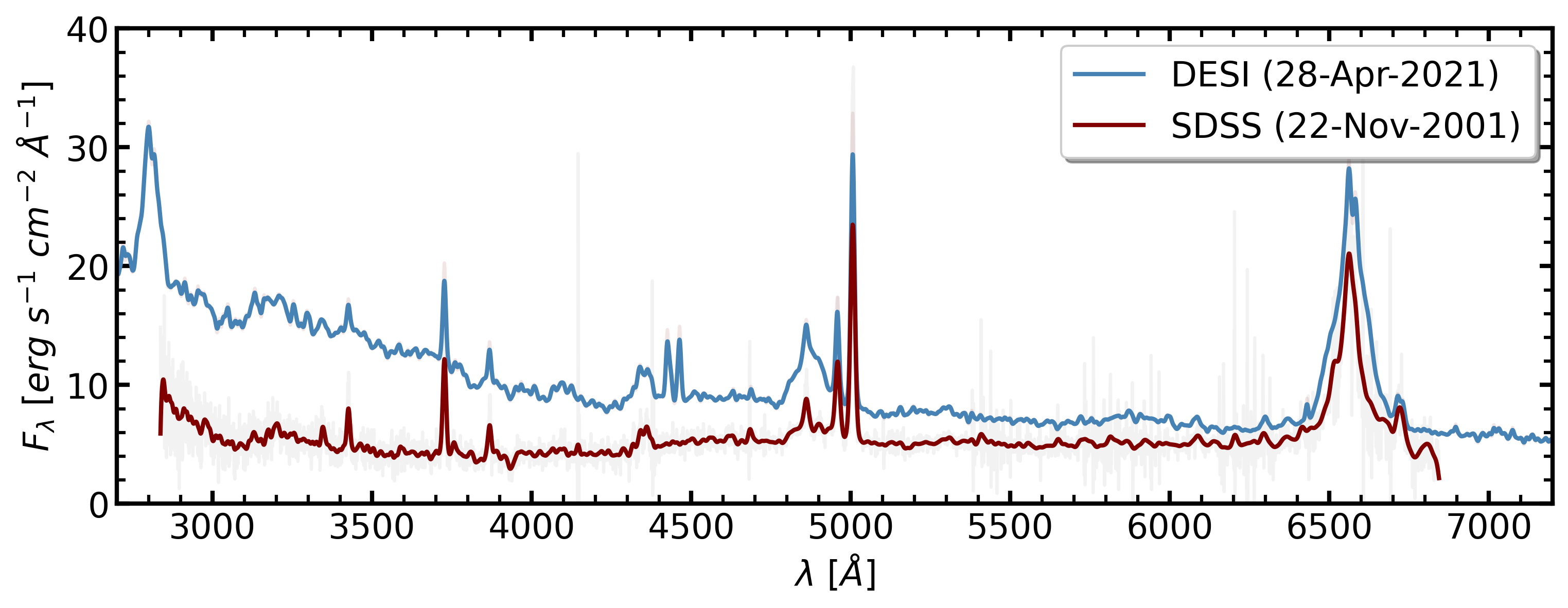}
    \caption{SDSS J095902.76+021906.3 (z=0.34579). \textit{Top panel}: Lasair LSST \textit{ugrizy} difference image light curve from the LSSTCam. The current Rubin/LSSTCam difference-photometry light curve spans 116 days, providing the first multi-band optical monitoring of this CL-AGN over a continuous Rubin observing season. The vertical dashed line marks the position of the latest epoch (14-June-2026, 23:44:08 UTC). Because the measured difference fluxes are negative, a conventional AB magnitude cannot be calculated directly from difference flux, \textit{f}. Here the magnitude scale represents the absolute difference-flux amplitude (see Eq. \ref{eq:diffmag}); \textit{Bottom panel}: Archival optical spectroscopy from SDSS (red) and DESI (blue) for this source. The foreground spectra are binned using a Gaussian kernel with a 5\AA\ binning. The unbinned spectra are plotted in the background (in grey). Notice the clear variations in the continuum as well as the emission lines between the two epochs separated by nearly 20 years, as reported in \citet{guo2025changing}.}
    \label{fig:case_study}
\end{figure*}

As a demonstration of the scientific potential of the CL-AGN database, we ingested the full catalog into the Rubin Alert brokers. In this work, we utilized the UK-based Lasair\footnote{\url{https://lasair.lsst.ac.uk/}}, in addition to NOIRLab's ANTARES\footnote{\url{https://antares.noirlab.edu/}}, Chile-based ALeRCE\footnote{\url{https://lsst.alerce.online/}}, and the France-Australian Fink\footnote{\url{https://lsst.fink-portal.org/}}, to identify known CL-AGN with ongoing Rubin observations and enable real-time monitoring of their photometric evolution. Most of the brokers provide a watchlist utility - where the user can upload their source list with names and coordinates, define a matching radius, and let the broker unravel any possible matches. The user can enable push notifications via email/Slack or other platforms to monitor and verify these alerts. Using a default association radius of 2 arcsec, 32 of the 1438 cataloged sources have generated Rubin alerts to date. Here we present the most promising case currently available: SDSS J095902.76+021906.3 ($z=0.34579$, see Figure \ref{fig:case_study}).\\

The Rubin light curve for this source presently spans MJD 61090.11--61205.99 (2026 February 19--June 14), corresponding to a temporal baseline of 115.9 days, after which the source became unobservable. During this interval, the source was observed 1202 times, comprising 54, 236, 237, 426, 235, and 14 epochs in the $u$, $g$, $r$, $i$, $z$, and $y$ bands, respectively. The photometric measurements shown in Figure~\ref{fig:case_study} were obtained from the Rubin difference-imaging photometry distributed through the Lasair alert broker using a custom watchlist developed for the CL-AGN catalog\footnote{\url{https://lasair.lsst.ac.uk/objects/170028500581351556/}}. The left-hand ordinate shows the calibrated Rubin difference flux ($f_{\rm diff}$; nJy), while the right-hand axis provides the corresponding difference magnitude computed from the absolute value of the difference flux,
\begin{equation}
m_{\rm diff}
=
31.4
-
2.5\log_{10}\!\left(\left|f_{\rm diff}\right|_{\rm nJy}\right),
\label{eq:diffmag}
\end{equation}
where the plotted branch corresponds to negative difference fluxes,
\begin{equation}
f_{\rm diff}
=
-
10^{(31.4-m_{\rm diff})/2.5}
\ {\rm nJy}.
\end{equation}
This secondary axis is included solely to provide an approximate photometric scale for the observed variability; unlike conventional apparent magnitudes, Rubin difference magnitudes describe the amplitude of the variable component relative to the reference image and should not be interpreted as the absolute brightness of the source. In addition, we highlight the archival optical spectroscopy from SDSS (red) and DESI (blue) for this source in the bottom panel of this figure. Notice the clear variations in the continuum as well as the emission lines between the two epochs separated by nearly 20 years, as reported in \citet{guo2025changing}.\\

\begin{figure}
    \centering
    \includegraphics[width=\linewidth]{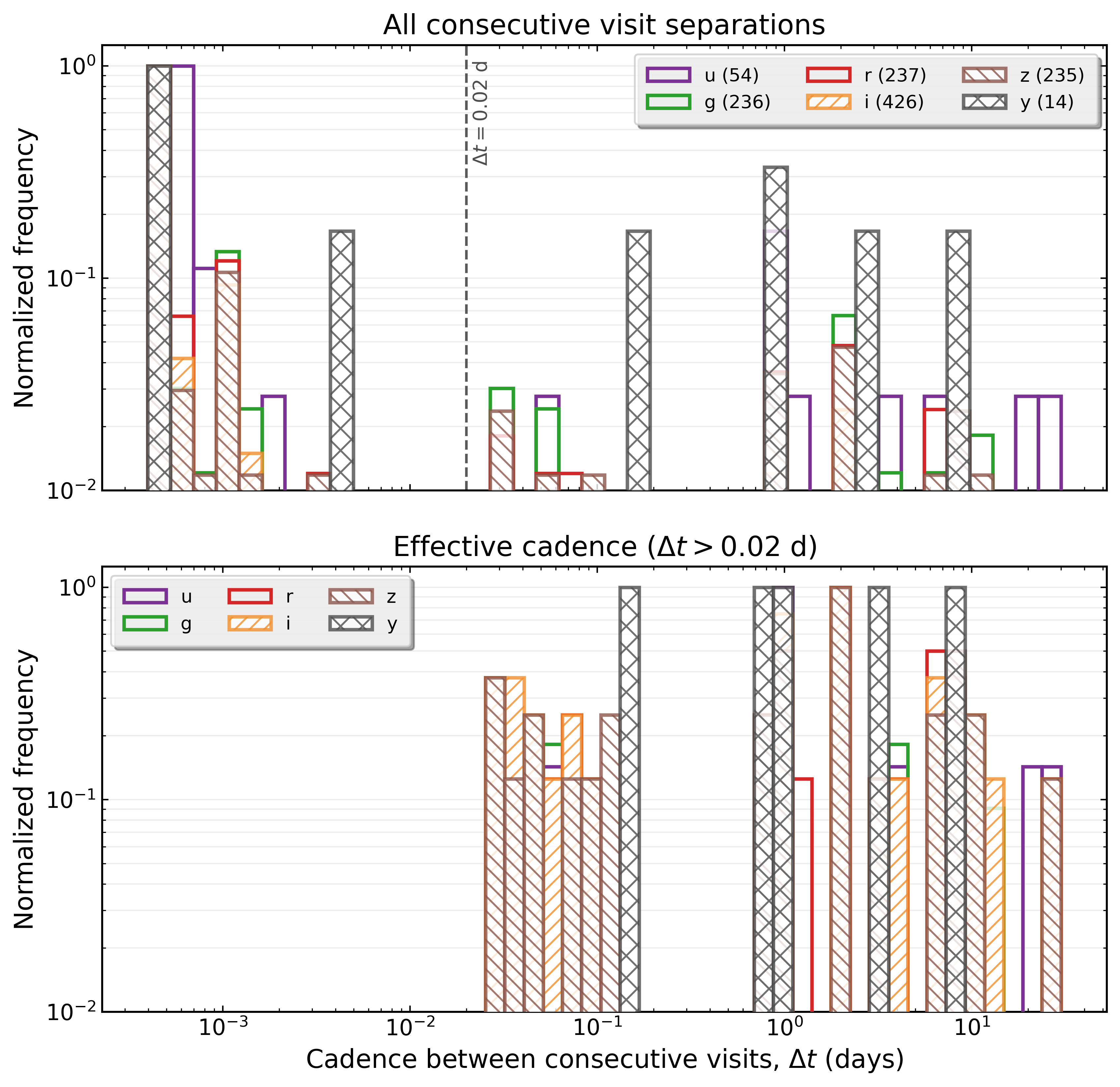}
    \caption{Distribution of consecutive visit separations for the Rubin/LSSTCam difference-photometry light curve. \textit{Top:} Histogram of all consecutive observations in each filter, including the closely spaced intra-night visit pairs characteristic of the Rubin observing strategy. The vertical dashed line marks the threshold of $\Delta t = 0.02$~d ($\approx29$~min) used to separate intra-night pairs from independent visits. \textit{Bottom:} Same distribution after excluding all separations shorter than 0.02~d, highlighting the effective sampling cadence relevant for the variability and reverberation analyses. Each histogram is normalized to its own maximum bin count to facilitate comparison of the cadence distributions among filters. The $griz$ bands provide the primary cadence for the monitoring campaign, whereas the $u$ and particularly the $y$ bands are substantially more sparsely sampled. The logarithmic axes emphasize the broad dynamic range of the cadence distribution, spanning from tens of seconds to several tens of days.}
    \label{fig:cadence}
\end{figure}

The observations comprise closely spaced intra-night visit pairs, typically separated by $\sim40$ s, together with an effective inter-night cadence of approximately one day in the $griz$ bands. As expected for an early survey dataset, weather losses and scheduling constraints introduce occasional gaps ranging from several days to a few weeks (see Figure \ref{fig:cadence}).\\

The Rubin difference-photometry light curves exhibit substantial variability in all six filters over the 115.9-day monitoring baseline. The peak-to-peak difference-flux amplitudes range from 10.56 to 13.63~$\mu$Jy, with comparable excursions across the well-sampled $griz$ bands (Table~\ref{tab:rubin_fluxstats}). The median photometric uncertainties are considerably smaller, ranging from 0.23~$\mu$Jy in the $g$ band to 0.64~$\mu$Jy in the $z$ band, corresponding to typical fractional uncertainties of $\lesssim2\%$ for the $griz$ filters. Consequently, the observed variability exceeds the median photometric uncertainty by factors of $\Delta f_{\rm pp}/\sigma_{\rm med}\sim20$--50 in the well-sampled bands, where $\Delta f_{\rm pp}$ is the peak-to-peak difference-flux variation and $\sigma_{\rm med}$ is the median photometric uncertainty. This demonstrates that the observed variability is robustly detected and is not dominated by photometric noise. The $u$ and particularly the $y$ bands exhibit larger median uncertainties (0.58 and 1.46~$\mu$Jy, respectively), reflecting their lower cadence and reduced sensitivity at the wavelength extremes. Although the $y$ band displays the largest peak-to-peak variation (13.63~$\mu$Jy), this measurement is based on only 14 epochs and should therefore be interpreted with caution.\\

\begin{table}[ht]
\centering
\footnotesize
\setlength{\tabcolsep}{4pt}
\caption{Summary statistics of the Rubin difference-photometry light curves. The peak-to-peak variation ($\Delta f_{\rm pp}$) is defined as the difference between the maximum and minimum measured difference flux in each filter, while $\sigma_{\rm med}$ denotes the median photometric uncertainty. The ratio $\Delta f_{\rm pp}/\sigma_{\rm med}$ provides a simple measure of the significance of the observed variability relative to the typical measurement error.}
\label{tab:rubin_fluxstats}
\begin{tabular}{lcccc}
\hline
Band & Epochs &
Median Error &
Peak-to-Peak &
Variability S/N \\
&
&
($\mu$Jy)
&
($\mu$Jy)
&
($\Delta f_{\rm pp}/\sigma_{\rm med}$)\\
\hline
$u$ & 54  & 0.580 & 11.22 & 19.3\\
$g$ & 236 & 0.231 & 11.70 & 50.6\\
$r$ & 237 & 0.298 & 11.15 & 37.4\\
$i$ & 426 & 0.363 & 10.56 & 29.1\\
$z$ & 235 & 0.640 & 13.54 & 21.1\\
$y$ & 14  & 1.456 & 13.63 & 9.4\\
\hline
\end{tabular}
\end{table}

The first Rubin LSSTCam data already provides an unprecedented six-band optical light curve for a spectroscopically confirmed changing-look AGN, enabling an initial investigation of its short-term time-domain behavior. We therefore carried out a suite of variability analyses, including characterization of the long-term chromatic evolution, searches for periodic variability using generalized Lomb--Scargle and Weighted Wavelet $Z$-transform periodograms, investigation of sampling aliases through the spectral window function, and inter-band continuum lag measurements using ICCF with FR/RSS uncertainty estimation. The principal results are summarized in Appendix~\ref{sec:timedomain}. In brief, the light curves exhibit coherent chromatic variability across the Rubin filters, but the current 115.9-day baseline and non-uniform cadence do not yet provide compelling evidence for statistically significant periodic variability or measurable inter-band continuum delays. Continued monitoring over subsequent Rubin observing seasons combined with improved templates from image stacking\footnote{\url{https://prompt-products.lsst.io/processing/templates/index.html}}, together with contemporaneous or quasi-simultaneous spectroscopy, will substantially improve sensitivity to both long-timescale variability and continuum reverberation.\\

\section{Summary and conclusions}
\label{sec:summary}

We have compiled and homogenized a catalog of known changing-look AGN and related candidate systems, and supplemented it with information from a broad range of multiwavelength surveys. Source names, coordinates, classifications, redshifts, and literature references were standardized into a common format. Positional matching and manual inspection were then used to remove duplicate entries while retaining the complete literature provenance of each object. The resulting optical parent catalog contains 1,438 unique sources, comprising 987 spectroscopically confirmed CL-AGN and 451 candidate or photometrically selected systems.\\

The optical sample was cross-matched with radio, infrared, ultraviolet, X-ray, millimeter, and gamma-ray resources. For the principal survey catalogs, the associations were evaluated with the Bayesian cross-matching code \texttt{nway} and survey-specific quality criteria. The resulting layers contain 188 sources with broad radio coverage, including 9 RFC/VLBI compact-core matches, as well as 16 sources with ALMA archival coverage. Infrared counterparts are available for 1,353 sources, GALEX ultraviolet counterparts for 918 sources, and X-ray counterparts for 576 sources. In addition, 9 candidate associations were found within the Fermi-LAT 95\% localization regions and are reported separately because of their greater positional uncertainty.\\

Overall, 1,385 sources, or 96.3\% of the optical parent sample, have at least one non-optical counterpart when the broad GALEX layer is included. Among them, 75 objects have coverage in all four of the main ancillary regimes considered here: radio, infrared, ultraviolet, and X-ray. This subset provides particularly valuable targets for studies that require information about several components of the AGN system, from accretion-disk and dust emission to high-energy and radio activity.\\

We also evaluated the overlap between the confirmed CL-AGN sample and the planned Rubin/LSST observing regions. Seventy spectroscopically confirmed sources lie within the Wide--Fast--Deep footprint, while five are located in the Deep Drilling Fields. These objects form an existing reference sample for investigating long-term variability with Rubin data and for connecting future optical light curves with archival observations at other wavelengths.\\

Rubin is expected to increase the known population of changing-look AGN by at least one to two orders of magnitude. Depending on the intrinsic transition rate and the available spectroscopic follow-up, the survey is expected to identify tens of thousands of photometric CL-AGN candidates and ultimately thousands to perhaps more than ten thousand spectroscopically confirmed changing-look AGN, transforming CL-AGN studies from small-sample investigations into population science. Recent systematic surveys indicate that approximately 1–6\% of nearby AGN undergo changing-look transitions over decade-long timescales, with measured lower limits of $>$6\% from the 6dFGS–SkyMapper survey \citep{hon2022skymapper} and 1.7–9.6\% from the 6dFGS–ATLAS survey \citep{amrutha2024discovering}. Extrapolating a conservative 1–3\% transition fraction to Rubin's expected sample of 20–50 million variable AGN suggests that Rubin LSST could identify roughly 2$\times$10$^5$- 1.5$\times$10$^6$ photometric changing-look AGN candidates during its 10-year survey, increasing the currently known population by more than two orders of magnitude. These discoveries will transform changing-look AGN studies from individual-object investigations into a statistical population science. Of these, we expect 10$^4$-10$^5$ spectroscopically confirmed CL-AGNs. Spectroscopic confirmation of Rubin changing-look AGN candidates will rely heavily on existing and forthcoming multiplexed spectroscopic surveys, including DESI-II \citep{DESI2022arXiv220903585S}, SDSS-V \citep{Kollmeier2017arXiv171103234K}, 4MOST \citep{4MOST2019Msngr.175....3D}, MOONS \citep{MOONS2020Msngr.180...10C}, and the planned Wide Field Spectroscopic Telescope \citep[WST,][]{WST2024arXiv240305398M}, complemented by rapid-response and targeted observations with facilities such as Gemini Observatory and the SOAR Telescope \citep[see e.g.,][]{GOATS2026arXiv260628645S}.\\

The catalog therefore establishes a unified reference sample for target selection, archival searches, and comparative studies of changing-look activity. By combining literature information with consistently defined multiwavelength and Rubin-footprint flags, it provides a foundation for studying the known CL-AGN population and for placing future time-domain discoveries in a broader observational context.


\begin{acknowledgments}
MC is grateful for the hospitality and to the staff at the International Gemini Observatory in La Serena, Chile. SP is supported by the International Gemini Observatory, a program of NSF NOIRLab, which is managed by the Association of Universities for Research in Astronomy (AURA) under a cooperative agreement with the U.S. National Science Foundation, on behalf of the Gemini partnership of Argentina, Brazil, Canada, Chile, the Republic of Korea, and the United States of America. 

\end{acknowledgments}

\facilities:{ ARC, ATLAS, ATT, Blanco, Bok, CTIO:1.5m, Du Pont, FTS, Hale, HET, ING:Herschel, KPNO:2.1m, LAMOST, LCOGT, Magellan:Baade, Magellan:Clay, Mayall, MMT, Perkins, PO:48, PS1, Skymapper, Sloan, UKST, VLT:Antu, VLT:Kueyen, Rubin:Simonyi, ALMA, VLA, ASKAP, MOST, LOFAR, WISE, 2MASS, AKARI, GALEX, ROSAT, eROSITA, XMM, CXO, Fermi}

\software{numpy \citep{numpy}, matplotlib \citep{matplotlib}, astropy \citep{2013A&A...558A..33A,2018AJ....156..123A}}, \texttt{astroquery} \citep{ginsburg2019astroquery}, 
and \texttt{nway} \citep{salvato2018finding}.

\begin{appendix}
\section{Time-domain analysis of the Rubin light curve for SDSS J095902.76+021906.3}
\label{sec:timedomain}

The Rubin/LSSTCam light curve provides an opportunity to investigate the short-term optical variability of this changing-look AGN using simultaneous six-band photometry. Our analysis proceeds by first characterizing the long-term variability before searching for periodic behavior and inter-band continuum delays. This sequence minimizes the impact of secular variability on subsequent timing analyses.\\

\subsection{Long-term chromatic variability}

The Rubin difference-photometry light curves span approximately 116 days and contain 54, 236, 237, 426, 235, and 14 measurements in the $u$, $g$, $r$, $i$, $z$, and $y$ bands, respectively. All bands exhibit coherent long-term variability, although the cadence varies substantially among filters.\\

To quantify the secular evolution, we fitted independent linear trends to each band. The well-sampled filters show slopes of $43.3$, $27.0$, $41.5$, and $-22.2$ nJy day$^{-1}$ in the $g$, $r$, $i$, and $z$ bands, respectively. The opposite sign measured in the $z$ band indicates that the variability is chromatic rather than a simple achromatic change in flux. However, the wavelength dependence is not monotonic, suggesting that the observed evolution cannot be described by a single varying blackbody or by a simple change in disk temperature. This is not unexpected, since the measured difference flux contains contributions from the variable continuum, broad emission lines, and the reference image, and therefore does not directly trace the intrinsic continuum spectral energy distribution \citep[e.g.,][]{Wilhite2005ApJ...633..638W,Schmidt2012ApJ...744..147S}.\\

\subsection{Search for periodic variability}

Periodic variability was investigated using generalized Lomb--Scargle (GLS) periodograms \citep{Zechmeister2009A&A...496..577Z} applied to the detrended light curves.\\ 

The strongest peak in the multiband periodogram occurs near 58 days. However, this corresponds closely to the maximum period explored in our search (approximately half the observing baseline), indicating that the periodogram is primarily tracing residual long-timescale variability. Additional peaks near 27, 16, and 12 days are present but are weaker and are not consistently recovered across all bands. Based on the current baseline, we find no evidence for a statistically robust periodic signal.\\

\subsection{Influence of the observing cadence}

To assess whether the candidate periods are related to the temporal sampling, we computed the spectral window function for each band \citep{Deeming1975Ap&SS..36..137D}.\\

The dominant window-function peaks occur near 50, 23--28, 16, and 5--8 days, closely matching the principal peaks identified by the GLS periodograms. In particular, the candidate periods near 27 and 16 days coincide with strong features in the sampling window (see Figure \ref{fig:cadence}), while the $\sim58$ day peak lies close to both the long-period window feature and the search boundary.\\

The close correspondence between the window-function structure and the GLS peaks suggests that the candidate periodicities are largely influenced by the observing cadence and should not be interpreted as evidence for intrinsic periodic variability.\\

\subsection{Time-frequency analysis}

To investigate whether the variability evolves with time, we computed Weighted Wavelet $Z$-transform (WWZ) periodograms \citep{Foster1996AJ....112.1709F}, which retain temporal information while searching for periodic signals.\\

The WWZ maps are dominated by power at the longest accessible timescales. Although transient enhancements appear near 25--30 days in several bands, these features are neither persistent throughout the observing window nor stable in frequency. Moreover, they coincide with the dominant aliases identified in the window-function analysis.\\

The WWZ analysis therefore supports the interpretation that the observed variability is dominated by long-term, non-periodic evolution.\\

\subsection{Inter-band continuum correlations}

We searched for wavelength-dependent continuum delays using the interpolated cross-correlation function (ICCF; \citealt{Gaskell1987ApJS...65....1G}; \citealt{White1994}). Lag uncertainties were estimated using the Flux Randomization/Random Subset Selection (FR/RSS) method \citep{1998PASP..110..660P, Peterson2004ApJ...613..682P}.\\

\begin{figure*}
    \centering
    \includegraphics[width=0.9\linewidth]{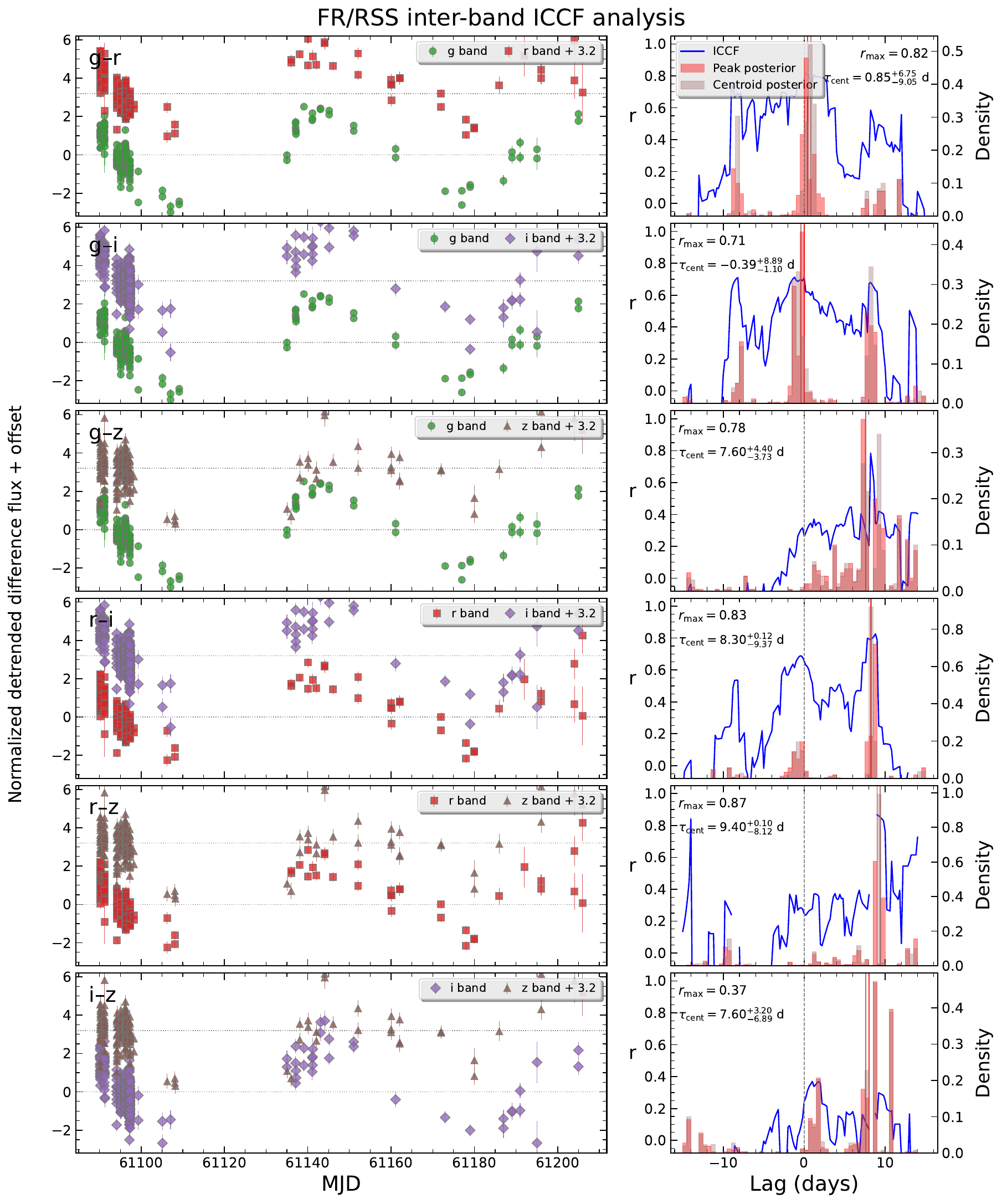}
    \caption{Flux Randomization/Random Subset Selection (FR/RSS) inter-band interpolation cross-correlation function (ICCF) analysis for the Rubin $g$, $r$, $i$, and $z$ light curves. For each band pair, the left panel shows the detrended, normalized difference-flux light curves (offset vertically for clarity), while the right panel shows the corresponding ICCF (blue curve) together with the FR/RSS posterior distributions of the peak (red) and centroid (brown) lags. The vertical lines indicate the median values of the peak and centroid lag distributions, and the panels list the maximum correlation coefficient ($r_{\rm max}$) and the median centroid lag with its 16th--84th percentile uncertainties. The optical bands exhibit strong correlations ($r_{\rm max}\approx0.73$--0.94), indicating coherent variability across the Rubin filters. However, the centroid-lag posteriors are generally broad and, in several cases, multimodal, with no self-consistent sequence of wavelength-dependent delays. The nominal long lags involving the $z$ band are comparable to half of the observational baseline and are therefore unlikely to represent physical continuum reverberation. The $u$ and $y$ bands were excluded from this analysis because of their sparse sampling (54 and 14 epochs, respectively), which is insufficient for reliable ICCF and FR/RSS lag estimation. Restricting the analysis to the well-sampled $g$, $r$, $i$, and $z$ light curves minimizes interpolation artifacts and improves the robustness of the inferred lag distributions.}
    \label{fig:ICCF}
\end{figure*}

The optical light curves are strongly correlated, with peak correlation coefficients ranging from $r_{\rm max}\approx0.73$ to 0.94. However, the FR/RSS lag distributions are generally broad and frequently multimodal (see Figure \ref{fig:ICCF}). For example, the $u-g$ centroid lag is

\begin{equation}
\tau_{ug}=0.19^{+1.58}_{-61.39}\ {\rm days},
\end{equation}

which is fully consistent with zero. Several band pairs involving the $z$ band yield centroid lags of approximately 60--64 days, comparable to half of the observational baseline. These values are substantially longer than expected for optical continuum reverberation and most likely reflect the limited baseline together with residual long-term variability.\\

Furthermore, the measured lag sequence is not internally consistent between different band pairs and does not satisfy the expected closure relations. We therefore do not find evidence for a robust detection of inter-band continuum lags.\\

Although the u-band contains 54 measurements, its cadence is substantially lower than that of the griz bands and contains large temporal gaps, resulting in poor overlap during ICCF interpolation and unstable FR/RSS lag distributions. We therefore restrict the analysis to the four well-sampled filters.\\

\subsection{Lag--wavelength relation}

The FR/RSS centroid lags referenced to the $u$ band were compared with the standard thin-disk prediction,

\begin{equation}
\tau(\lambda)\propto\lambda^{4/3},
\end{equation}

expected for continuum reverberation in a geometrically thin, optically thick accretion disk \citep{SS1973A&A....24..337S, Collier1999MNRAS.302L..24C, Cackett2007MNRAS.380..669C}. A weighted fit yields

\begin{equation}
\tau_0=0.90\pm1.06~{\rm days},
\end{equation}

consistent with zero. Although the $z$-band measurement lies close to the fitted relation, the remaining bands exhibit large asymmetric uncertainties arising from broad FR/RSS posterior distributions. Consequently, the present data do not provide a statistically significant detection of the expected wavelength-dependent continuum lag.\\

In summary, the Rubin light curve clearly reveals coherent, chromatic variability over the first 116 days of monitoring. After accounting for the long-term trend, however, we find no convincing evidence for periodic variability or measurable inter-band continuum lags. The strongest candidate periods coincide with prominent features in the sampling window, while the lag posteriors remain broad and, in several cases, multimodal. These results indicate that the current temporal baseline is sufficient to characterize the secular photometric evolution, but is not yet long enough to place strong constraints on periodic variability or continuum reverberation. Continued monitoring over multiple Rubin observing seasons will substantially improve sensitivity to long-timescale variability and inter-band delays. Combining these photometric data with contemporaneous or quasi-simultaneous spectroscopy will also enable direct comparisons between the photometric evolution and changes in the broad emission lines, providing a more complete picture of the changing-look phenomenon.

\end{appendix}

\bibliography{main}{}
\bibliographystyle{aasjournalv7}

\end{document}